\documentclass[authoryear,3p]{elsarticle}

\usepackage{amssymb,amsmath,amsfonts,amstext}
\usepackage{graphicx}
\usepackage{subfigure}
\usepackage{mathtools}
\usepackage{stmaryrd}
\usepackage{color}

\journal{Journal of the Mechanics and Physics of Solids}

\begin{document} 

\begin{frontmatter}

\title{Kinematic description of crystal plasticity in the finite kinematic framework: a micromechanical understanding of $F=F^e F^p$}

\author[label1,label2,label3]{C.~Reina\corref{cor1}}
\author[label2]{S.~Conti}
\address[label1]{Lawrence Livermore National Laboratory, Livermore, CA 94550, United States}
\address[label2]{Institut fur Angewandte Mathematik, Universitat Bonn, 53115 Bonn, Germany}
\address[label3]{University of Pennsylvania, Philadelphia, PA 19104-6315, United States}
\cortext[cor1]{creina@seas.upenn.edu}

\begin{abstract}
The plastic component of
the deformation gradient plays a central role in finite kinematic
models of plasticity. However, its characterization has been the source of
extended debates in the literature and many important issues still remain
unresolved. Some examples are the micromechanical understanding of
$\mathbf{F}=\mathbf{F}^e\mathbf{F}^p$ with multiple active slip systems, the
uniqueness of the decomposition, or the characterization of the plastic
deformation without reference to the so-called intermediate configuration. In
this paper, we shed some light to these issues via a two-dimensional kinematic
analysis of the plastic deformation induced by discrete slip surfaces and the
corresponding dislocation structures. In particular, we supply definitions for
the elastic and plastic components of
the deformation gradient as a function of the active slip
systems without any a priori assumption on the decomposition of the total
deformation gradient. These definitions are explicitly and uniquely given from
the microstructure and do not make use of any unrealizable intermediate
configuration. The analysis starts from a semi-continuous mathematical
description of the deformation at the microscale, where the displacements are 
considered continuous everywhere in the domain except at the discrete slip
surfaces, over which there is a displacement jump. At this scale, where the
microstructure is resolved, the deformation is uniquely characterized from
purely kinematic considerations and the elastic and plastic components of
the deformation
gradient can be defined based on physical arguments. These quantities are then
passed to the continuous limit via  homogenization, i.e., by increasing the
number of slip surfaces to infinity and reducing the lattice parameter to
zero. This continuum limit is computed for several illustrative examples,
where the well-known multiplicative decomposition of the total deformation
gradient is recovered. Additionally, by similar arguments, an expression of
the dislocation density tensor is obtained as the limit of discrete
dislocation densities which are well characterized within the semi-continuous
model. 
\end{abstract}


\end{frontmatter}

\section{Introduction}
Experiments have shown that the elastic behavior of polycrystalline materials is fairly insensitive to prior plastic distortion. It has then become convenient for the formulation of constitutive theories to separate the total macroscopic deformation into an elastic part and a plastic one. This decomposition is commonly done additively ($\boldsymbol \varepsilon = \boldsymbol \varepsilon^e + \boldsymbol \varepsilon^p$) in the linearized kinematic framework and multiplicatively ($\mathbf{F} = \mathbf{F}^e\mathbf{F}^p$) in finite deformations.

The elastic and plastic deformation tensors ($\boldsymbol \varepsilon^e, \mathbf{F}^e$ and $\boldsymbol \varepsilon^p, \mathbf{F}^p$) physically measure the effective shape change induced by elastic and plastic mechanisms respectively. However, their contribution to the total deformation of the body cannot generally be visualized via intermediate realizable configurations where only one of the effects is considered. For crystalline materials, this is physically due to the presence of dislocations, which induces kinematic constraints on the elastic deformation of the material surrounding them. Despite the aforementioned coupling, the additive decomposition in the linearized setting ($\boldsymbol \varepsilon = \boldsymbol \varepsilon^e + \boldsymbol \varepsilon^p$) can be easily justified from the underlying kinematics of slip via the principle of superposition and simple convergence rules. In finite deformations though, the problem is far more complex and several decompositions have been proposed in the literature: some additive \citep{NematNasser1979, Zbib1993, Pantelides1994, Shen1998} and others multiplicative ($\mathbf{F}=\mathbf{F}^p\mathbf{F}^e$ \citep{Clifton1972, Lubarda1999}, $\mathbf{F}=\mathbf{F}^e\mathbf{F}^p$  \citep{LeeLiu1967}). The expression of the form $\mathbf{F}=\mathbf{F}^e\mathbf{F}^p$ has become standard in continuum models of finite elastoplasticity. However, several issues related to this decomposition are still topics of ongoing discussion. Some important examples are:

\begin{enumerate}
\item[$\cdot$] The conditions for the existence of the decomposition
  $\mathbf{F}=\mathbf{F}^e\mathbf{F}^p$\citep{CaseyNaghdi1992, Owen2002, DelpieroOwenBook2004}.
\item[$\cdot$] The uniqueness of the decomposition \citep{NematNasser1979,
    LubardaLee1981, Zbib1993, Naghdi1990, GreenNaghdi1971, CaseyNaghdi1980,
    Mandel1973, Dafalias1987, Rice1971} and the associated invariance
  requirements on elastoplastic constitutive laws \citep{Mandel1973,
    VanDerGiessen1991, Dafalias1998, ClaytonMcDowell2003}. 
\item[$\cdot$] The understanding of the  plastic deformation tensor
  $\mathbf{F}^p$ without reference to an artificial intermediate configuration
  and its direct relation to the underlying dislocation structure and active
  slip systems  \citep{DelpieroOwenBook2004}.
\item[$\cdot$] The appropriate measures for the dislocation content in a body as function of the elastic or plastic deformation tensor \citep{Bilby1955, Eshelby1956, Kroner1960, Fox1966, Willis1967, AcharyaBassani2000, CermelliGurtin2001}.
\end{enumerate}

A deep understanding of this subject is essential for the formulation of
physically sound constitutive equations and multiscale models of material
behavior
\citep{OrtizRepetto1999,DeseriOwen2000, DeseriOwen2002, CermelliLeoni2005,GarronileoniPonsiglione2010,Roters2010,ScardiaZeppieri2012}. The
goal of this work is precisely to take a step in that direction and shed some
light to the aforementioned issues.

The main idea behind this work lies on the fact that kinematic considerations
are sufficient to characterize the elastic and plastic deformation of the
material if the microstructure (dislocations and slip planes) is fully
resolved. This is in contrast with the standard macroscopic perspective, where
energetic considerations are required in order to characterize the deformation
as elastic or plastic. 
The theory of structured deformations provides some ideas towards this
direction, including in particular working with a two-scale
geometry and providing an approximation  theorem based on sequences of
piecewise classical deformations exhibiting 
jumps (slips for the case of plasticity) across disarrangement sites, see
 \citep{DelpieroOwen1993}, \citep{DelpieroOwenBook2004} and the discussion in
Section \ref{secpreviousstudies} below.

We perform in this paper a
thorough kinematic analysis of elastoplastic deformations at the
microscale, building upon a full kinematic resolution of the 
slip planes. This study will provide physical definitions of the deformation
tensors $\mathbf{F},\mathbf{F}^e,\mathbf{F}^p$ that are `unique', static
(not in rate form) and that do not make use of any intermediate
configuration.

The paper is organized as follows. We begin in Section \ref{Sec:Continuum} by summarizing the standard description of finite elastoplasticity at the continuum scale and discuss several measures of dislocation densities. In Section 3, we take a careful look at the microstructures induced by dislocation glide and describe the resulting deformations via a semi-continuous mapping that accounts for the discontinuity at the slip planes while remaining continuous in the remainder of the domain. This formulation is applied in Section \ref{Sec:Examples} to several examples of elastoplastic deformations, from which appropriate definitions of the elastic and plastic deformation tensors at the discrete level are inferred. These microscopic definitions are then passed to the continuum scale via mathematical homogenization under suitable scaling, recovering the well-known decomposition $\mathbf{F}=\mathbf{F}^e\mathbf{F}^p$. In Section 5, a similar analysis is performed for the dislocation structures, and physical definitions of the dislocation density tensor at the discrete and continuum scale are provided as a function of the previously obtained plastic deformation tensor. Based on these analyses and the acquired micromechanical understanding of crystal finite elastoplasticity, the previously enumerated issues are discussed in Section \ref{Sec::Additional}. Finally, the paper is concluded with Section \ref{Sec::Conclusion}, where the main results are summarized.


\section{Continuum description of elastoplasticity and the dislocation density tensor in finite kinematics} \label{Sec:Continuum}
In continuum solid mechanics, a Lagrangian formulation is typically adopted to
describe the deformation of a body. From this perspective, the final
configuration of each material point $\mathbf{X}$ in a body $\Omega \subset
\mathbb{R}^d$ ($d=2,3$) is described via a continuous deformation mapping:
$\mathbf{x} = \boldsymbol \varphi( \mathbf{X})$. The deformation gradient
$\mathbf{F}$ is then defined as $F_{iJ}=\frac{\partial \varphi_i}{\partial
  X_J}$, which under sufficient smoothness assumptions, satisfies $\text{Curl
} \mathbf{F} = 0$ \footnote{$\text{Curl}\ \mathbf{F}$ is
    defined in this paper as $(\text{Curl}\ F)_{iJ} = F_{iM,K} e_{JKM}$ in
    three dimensions (d=3) and $(\text{Curl}\ F)_i = F_{iM,K} e_{3KM}$ in two
    dimensions (d=2), where the derivatives are performed with respect to the
    coordinates in the reference configuration. Note that these definitions
  might differ from that of other authors. }. Additionally, the condition of
non-interpenetration of matter requires that $\boldsymbol \varphi(
\mathbf{X})$ is one-to-one and that $\det \mathbf{F} > 0$. 

This continuum description of the deformation does not provide any information on how the final configuration is realized: if elastically or by means of other inelastic mechanisms. This decoupling is essential in the formulation of constitutive relations and is generally described by means of an elastoplastic decomposition of the deformation gradient of the type $\mathbf{F}=\mathbf{F}^e\mathbf{F}^p$ \citep{LeeLiu1967}. In this paper we will only consider inelasticity induced by dislocation glide and disregard other processes such as phase transformations, diffusion, void nucleation or appearance of microcracks. The plastic deformation is then restricted to be volume preserving, i.e. $\det \mathbf{F}^p =1$, and the elastic energy has then a unique minimum at $\mathbf{C}=\mathbf{F}^T\mathbf{F} =\mathbf{I}$, when the reference configuration is taken to be a perfect crystalline structure.
 
\begin{figure}
\begin{center}
    {\includegraphics[width=0.6\textwidth]{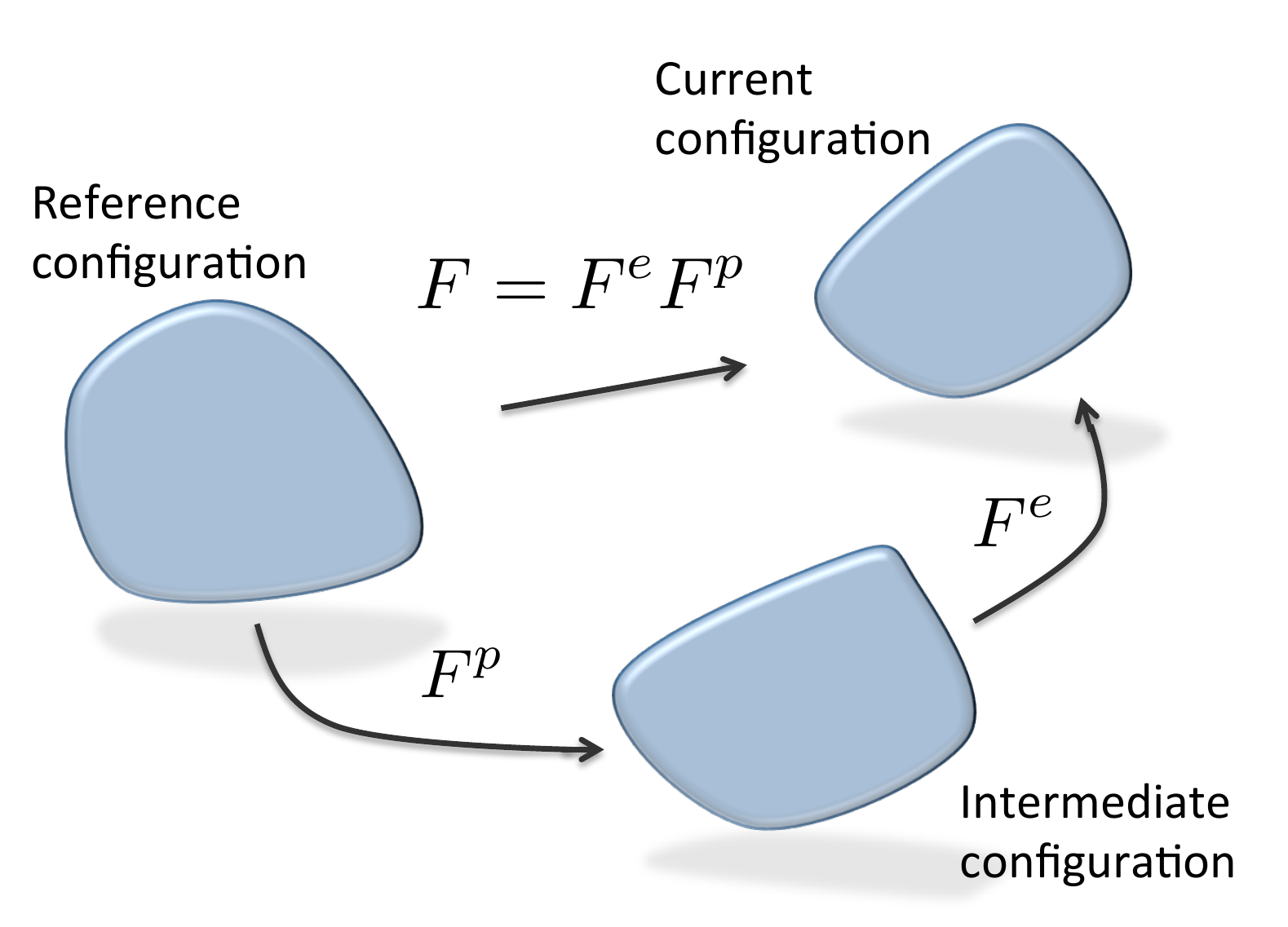}}
    \caption[]{Standard view of the decomposition of the total deformation gradient $\mathbf{F}$ into an elastic and plastic part ($\mathbf{F}^e$ and $\mathbf{F}^p$ respectively).}
    \label{Fig:FeFp}
\end{center}
\end{figure}

The elastoplastic decomposition in its classical understanding makes use of an
intermediate configuration in which the material is only deformed plastically,
cf. Fig. \ref{Fig:FeFp}. 
The configurations displayed in 
Figure  \ref{Fig:FeFp} are understood to be local, in particular the
intermediate one displays only a neighbourhood of points and not the whole
region of space occupied by the body.
As indicated in the introduction, such configuration 
is in general fictitious, since $\mathbf{F}^p$ does not always derive from a
deformation mapping (or equivalently, it does not satisfy the compatibility
conditions). It is typically seen as an ensemble of incompatible
  differential domains which are not uniquely determined by  $\mathbf{F}$. This non-uniqueness is removed in crystal plasticity by relating the plastic deformation to the underlying slip activity via  \citep{Rice1971}

\begin{equation} \label{Eq:FpRate}
\dot{\mathbf{F}}^p \mathbf{F}^{p-1} = \sum_k \dot{\gamma}_k \mathbf{M}_k \otimes \mathbf{N}_k,
\end{equation}
where $\dot{\gamma}_k$ is the strain rate on direction $\mathbf{M}_k$ of a slip system with normal $\mathbf{N}_k$.

\subsection{Compatibility and dislocation density tensor}
A tensor field $\mathbf{F}$ is said to be compatible over a domain $\Omega$ if
there exists a deformation mapping $\boldsymbol \varphi$ such that $\mathbf{F}
= \nabla \boldsymbol \varphi$. If the domain $\Omega$ is simply-connected,
$\mathbf{F}$ is compatible if and only if its \textit{Curl} vanishes
identically over the domain. As previously mentioned, the macroscopic
deformation gradient is by definition compatible, whereas the elastic and
plastic part do not have in general a vanishing $\textit{Curl}$. Their
incompatibilities are related through the expression $\mathbf{F}
=\mathbf{F}^e\mathbf{F}^p$ and satisfy  
\begin{equation}
\begin{split}
&\text{Curl}\ \mathbf{F}^{p} = \det (\mathbf{F})\ (\text{curl}\ \mathbf{F}^{e-1}) \mathbf{F}^{-T}, \quad \text{in 3D} \\
&\text{Curl}\ \mathbf{F}^p = \det (\mathbf{F})\ \text{curl}\ \mathbf{F}^{e-1} , \quad \text{in 2D}, \\
\end{split}
\end{equation}
where $\textit{curl}$ represents the rotational with respect to the coordinates in the deformed configuration. The proof for these expressions is detailed in \ref{Sec::AppendixA}.

The incompatibilities of $\mathbf{F}^p$ (or $\mathbf{F}^e$) in the context of crystal plasticity are physically induced by the presence of dislocations\footnote{Other defects can induce incompatibilities in the system. However, only dislocations will be considered in this work}, and are related mathematically via the so called dislocation density tensors $\mathbf{G}$. More specifically, $\mathbf{GN}$ measures the net Burgers vector per unit area of the dislocations piercing an infinitesimal surface normal to $\mathbf{N}$ \footnote{Other authors use the convention $\mathbf{G}^{\bot}\mathbf{N}$ to define the dislocation density tensor.}. Several expressions for $\mathbf{G}$ have been proposed in the literature based, primarily, on arguments in the setting of non-Riemannian geometry or by reasoning with Burgers circuits in the different configurations \citep{CermelliGurtin2001, AcharyaBassani2000, Acharya2004, Willis1967}. Although the majority of these expressions are equivalent when referred to the same reference frame \citep{Steinmann1996}, no complete agreement exists on its definition. Indeed, two commonly used, yet distinct, dislocation density tensors, where both the Burgers vector and normal are measured in the reference configuration, are
\begin{equation}
\mathbf{G}=\text{Curl}\ \mathbf{F}^p, \quad \mathbf{G}=\mathbf{F}^{p-1} \text{Curl}\ \mathbf{F}^p.
\end{equation}

In all cases, the $Curl$ is the main operator used to characterize the net content of dislocations. An important question may then rise when considering several elastoplastic deformations. Without loss of generality, consider two sequential deformations $\mathbf{F}^{p}_1$ and $\mathbf{F}^{p}_2$ and assume that the total plastic deformation can be expressed as $\mathbf{F}^{p}_2 \mathbf{F}^{p}_1$. The total amount of dislocations in the body, expressed as $\text{Curl}\ (\mathbf{F}^{p}_2 \mathbf{F}^{p}_1)$, reads
\begin{equation} \label{Eq:Curl_on_product}
\begin{split}
\left(\text{Curl}(F^{p}_2 F^{p}_1) \right)_{IJ} &= (F^{p}_{2IN}F^{p}_{1NM})_{,K} e_{JKM} \\
&= F^{p}_{2IN} (\text{Curl} F^{p}_1)_{NJ} + F^{p}_{1NM} F^{p}_{2IN,K} e_{JKM},
\end{split}
\end{equation}
 which is not a function of $\text{Curl}\ \mathbf{F}^{p}_1$ and
  $\text{Curl}\ \mathbf{F}^{p}_2$ alone. In the remainder of the paper we will provide a micromechanical understanding of the functional form of Eq.~(\ref{Eq:Curl_on_product}) attendant to the dislocations formed at each stage of the deformation.

\section{Semi-continuous description plastic activity induced by dislocation glide} \label{Sec:SemiContinuousModel}
\begin{figure}[ht]
\centering
\subfigure{
 {\includegraphics[trim=1cm 3cm 1cm 1cm, clip=true, width=0.45\textwidth]{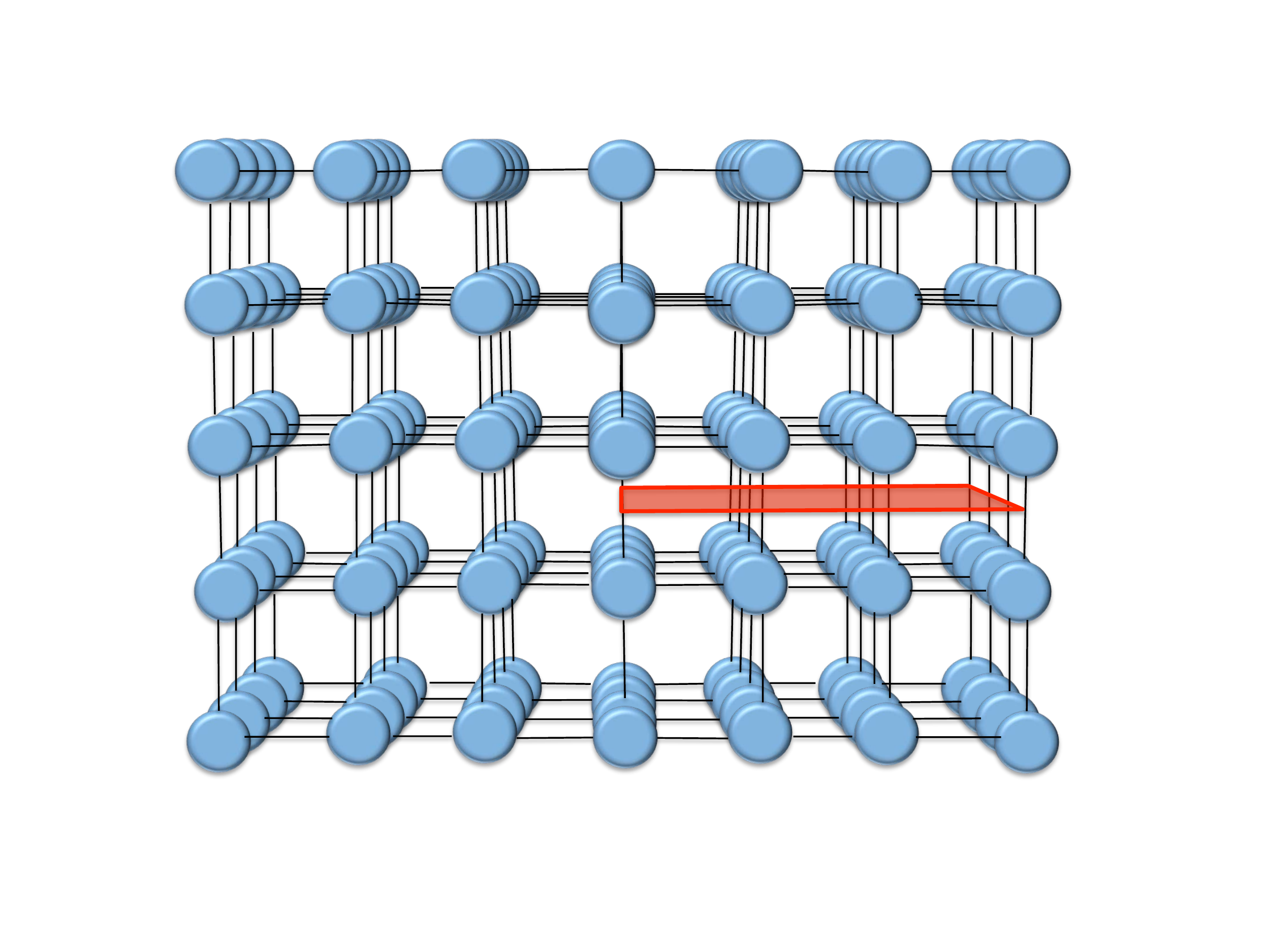}}}
\subfigure{
  {\includegraphics[trim=1cm 3cm 1cm 1cm, clip=true, width=0.45\textwidth]{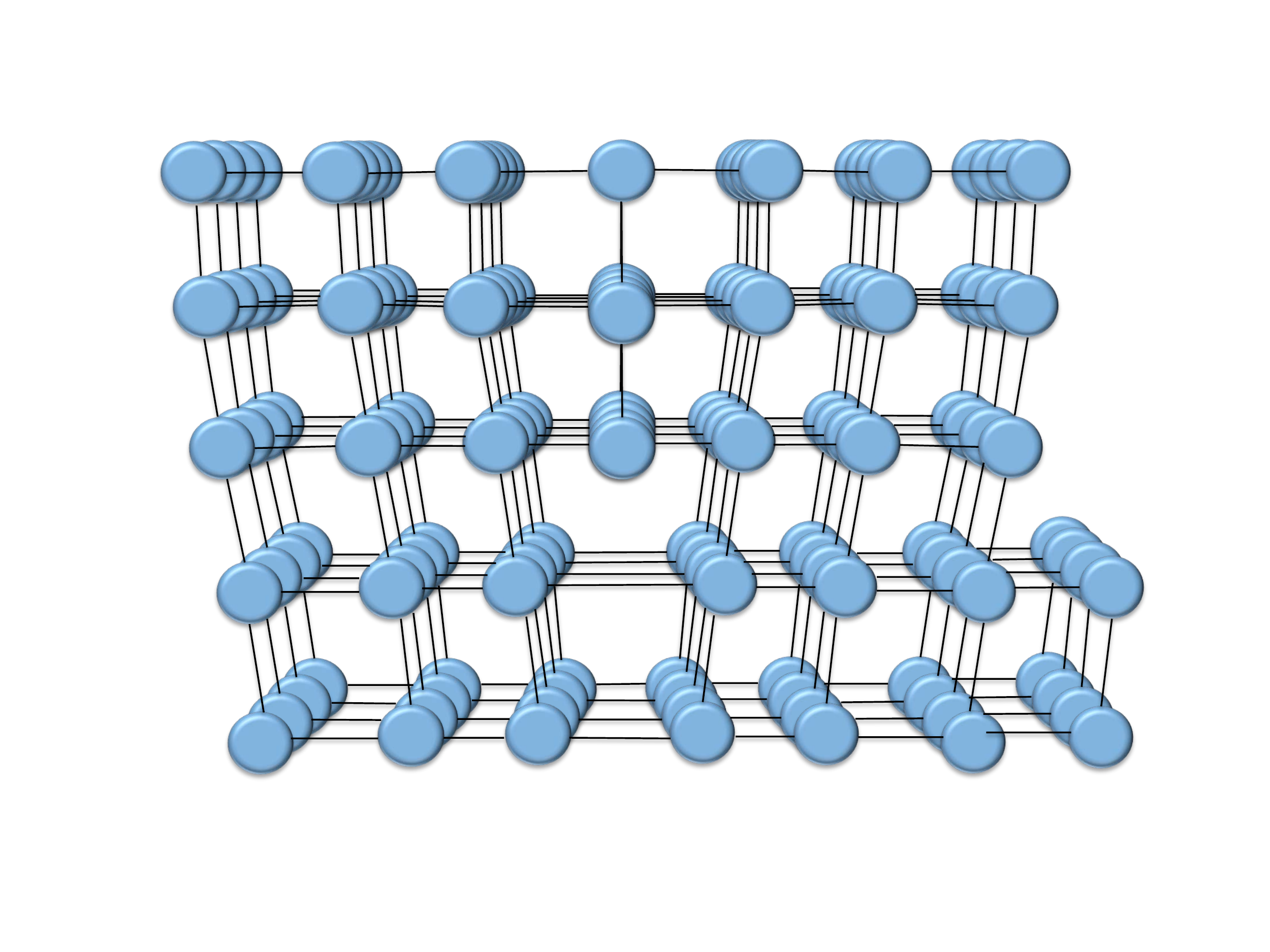}}}
\caption{Plastic deformation (right image) of a perfect three dimensional crystal (left image). The surface of displacement discontinuity is represented in the undeformed configuration.}
\label{Fig:Slip3D}
\end{figure}

From a physical perspective, the plastic deformation in crystalline materials is induced by the motion of dislocations, leaving behind surfaces of displacement discontinuity when referred to a perfect crystal, cf. Fig.~\ref{Fig:Slip3D}. Mathematically, we characterize these deformations with a Lagrangian semi-continuous model that tracks these surfaces of discontinuity while describing the material as continuous in the remainder of the domain. For the sake of simplicity, we initially restrict the model to a two-dimensional space, where slip occurs over lines and the dislocations (of edge type) are the endpoints of these lines. Extensions of these ideas to the three dimensional setting are discussed later on in Sec.~\ref{Sec:3D}. Similarly, the reference configuration for the Lagrangian description will be considered a perfect crystal.

\begin{figure}
\begin{center}
    {\includegraphics[width=0.55\textwidth]{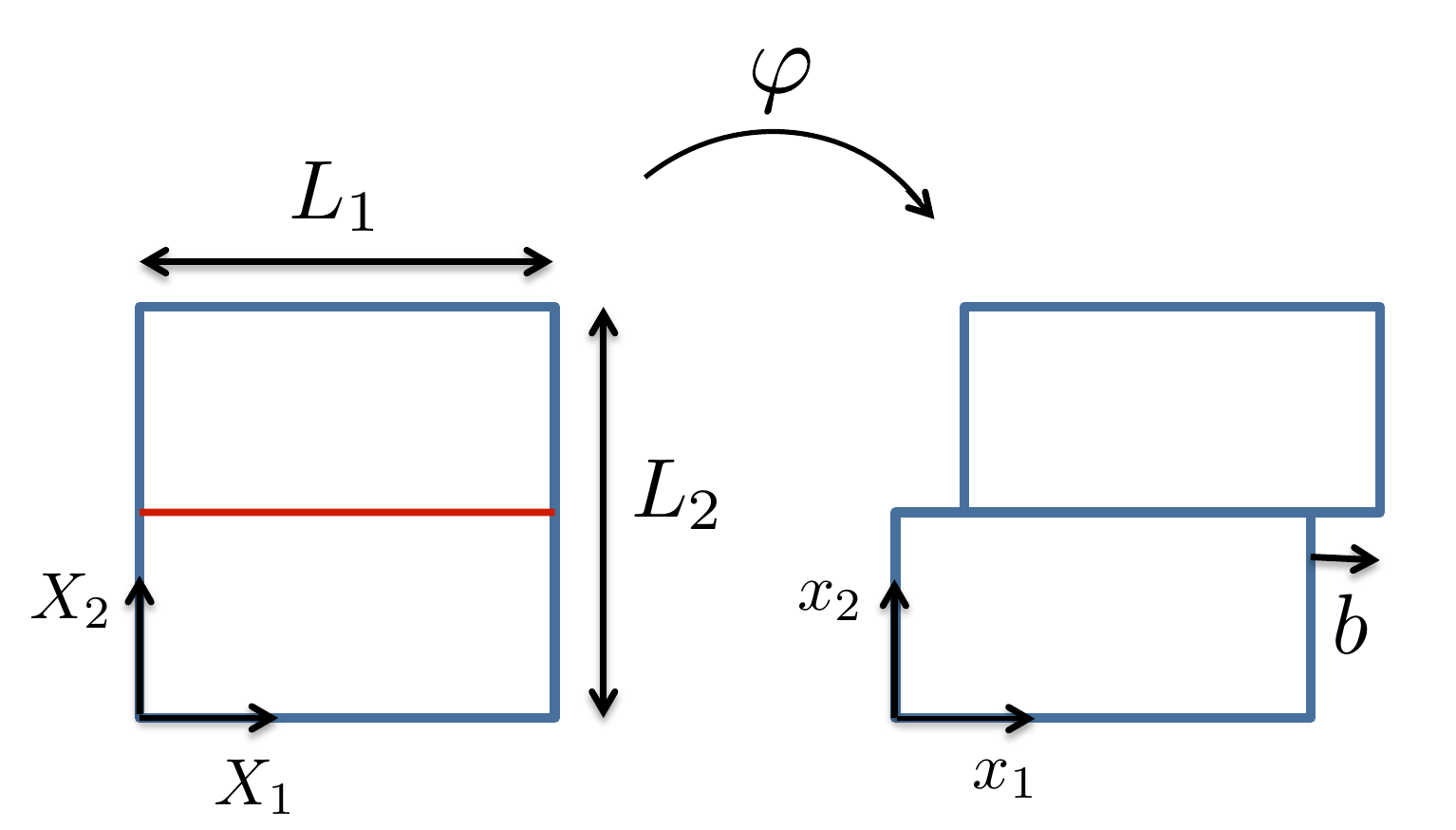}}
    \caption[]{Deformation (right image) induced by slip on a single jump set, represented in red in the reference configuration (left image). The amount of slip is given by the Burgers vector $\mathbf{b}$.}
    \label{Fig:Case1a}
\end{center}
\end{figure}

An example of the type of deformations here considered is shown in Fig. \ref{Fig:Case1a}, and is characterized by the following deformation mapping
\[ \mathbf{x}  = \boldsymbol \varphi (\mathbf{X})= \left \{ \begin{array} {l l }
\mathbf{X} & \text{if } X_2 < \frac{L_2}{2} \\
\mathbf{X} + \mathbf{b} & \text{if } X_2 > \frac{L_2}{2}. \\
\end{array} \right.
\]

This type of functions belong to the space of special functions of bounded
variation or SBV\footnote{At this scale, where dislocations and slip lines are
  resolved, the amount of slip and its support are both physically bounded. It
  is then justified to use special functions of bounded variation to describe
  the micromechanics of elastoplastic deformations.},
cf. \citep{EvansGariepy,AmbrosioFuscoPallara2000}, where we further require that
$\boldsymbol \varphi$ is one-to-one outside of the lines of discontinuity in order to avoid interpenetration of matter. The deformation gradient $\mathbf{F}$ can now be defined as its distributional gradient (generalization of the concept of derivative for non-differentiable functions), which for SBV functions is a measure of the form 
\begin{equation} \label{Eq:Definition_F}
\mathbf{F} = D \boldsymbol \varphi = \nabla \boldsymbol \varphi\ \mathcal{L}^2 + \llbracket \boldsymbol \varphi \rrbracket \otimes \mathbf{N}\ \mathcal{H}^1 \lfloor_{\mathcal{J}}. 
\end{equation}

The measure $\mathbf{F}$ consists of an absolutely continuous part defined
almost everywhere ($\nabla \boldsymbol \varphi\
\mathcal{L}^2$)\footnote{$\mathcal{L}^2$ is the Lebesgue measure in
  $\mathbb{R}^2$. It coincides with the standard measure of area in the
  plane.} and a singular part  that has its support over the slip lines or
jump set $\mathcal{J}$ ($ \llbracket \boldsymbol \varphi \rrbracket\otimes
\mathbf{N}\ \mathcal{H}^1 \lfloor_{\mathcal{J}} $)\footnote{$\mathcal{H}^1
  \lfloor_{\mathcal{J}}$ refers to the one-dimensional Hausdorff measure
  restricted to the jumpset $\mathcal{J}$. It delivers the length of a segment
  if contained in $\mathcal{J}$ and zero otherwise.}. 
$\nabla \boldsymbol \varphi$ is called the approximate
differential of $\boldsymbol \varphi$ and  corresponds to the `standard'
gradient where the function is differentiable (in particular, only outside of
the jump set). By definition, $\mathbf{F}$ has a vanishing \textit{Curl}. However, Curl $\nabla \boldsymbol \varphi$ does not vanish in general, and the same occurs for the singular part of the deformation gradient.

The regular part $\nabla \boldsymbol \varphi\ \mathcal{L}^2$ corresponds to the total deformation $\mathbf{F}$ at the points where no slip occurs and it can therefore be physically identified with the elastic deformation tensor
\begin{equation}\label{Eq:Definition_Fe}
\mathbf{F}^e \vcentcolon = \nabla \boldsymbol \varphi.
\end{equation} 

\begin{figure}
\begin{center}
    {\includegraphics[width=0.3\textwidth]{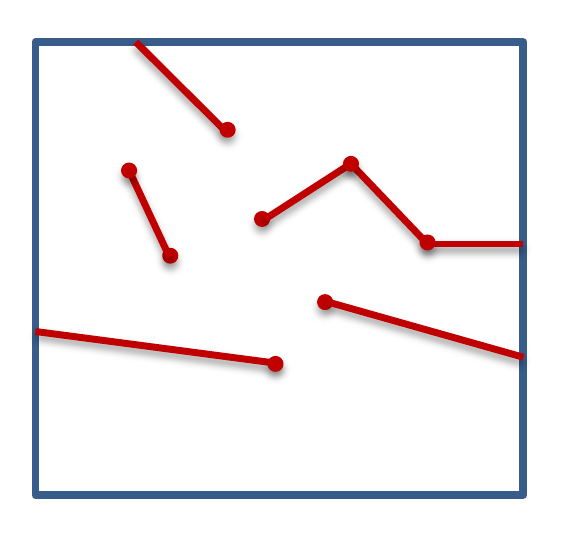}}
    \caption[]{Potential configuration for non-intersecting slip lines on a perfect crystal. They are required to be the union of straight segments and to terminate at dislocation points or at the boundary of the domain. The set of admissible slip line orientations depends on the specific crystallographic structure.}
    \label{Fig:SlipLines}
\end{center}
\end{figure}

For its part, $\llbracket \boldsymbol \varphi \rrbracket (\mathbf{X})
\vcentcolon= \boldsymbol \varphi^+(\mathbf{X}) - \boldsymbol
\varphi^-(\mathbf{X})$ is the displacement jump at point $\mathbf{X}$ in the
slip line, where the $+$ side is the one indicated by the normal $\mathbf{N}$
to the line at point $\mathbf{X}$. Without further restrictions on $\llbracket
\boldsymbol \varphi \rrbracket$, the class of deformation mappings considered
thus-far can also model processes such as cavitation, crack opening or
interpenetration of matter. We therefore proceed to analyze the kinematics of
slip to reduce the space of $\boldsymbol \varphi$ to that of elastoplastic
deformations induced by dislocation glide. We begin by identifying the
possible shapes of the slip lines when they do not intersect with each
other. Towards that goal, consider a simply connected subset
  $\omega$ of the domain $\Omega$ free of dislocations and traversed by a
  single slip line. Since $\omega$ is dislocation free, the dislacement jump
  recovers 
  a material free of defects in it\footnote{We are considering
      perfect dislocations, 
      or partials with negligible stacking fault energy.}. As a result, it is
  possible to envisage a plastic deformation in $\omega$ with no elastic
  energy\footnote{The elastic energy is defined up to a constant
      and is considered to be null here for the perfect crystalline structure
      in the reference configuration.}, or equivalently, with $\mathbf{F}^e =
  \nabla \boldsymbol\varphi\ \in SO(2)$. Such realization, by compatibility,
  leads to regions of homogeneous deformation \citep{Gurtin}, i.e. with
  constant rotation, at each side of the jump set. This condition combined
with the fact that no opening or interpenetration of matter is
  allowed for reduces the potential jump set in the reference configuration to
  a segment of a circle or of a straight line. From these two possibilities,
  only the later allows for small plastic distortions compatible with the
  underlying translational periodicity of crystalline materials. As a result
  we can conclude that non-intersecting jump sets in two dimensions consist
of an ensemble of straight segments in the reference configuration that
terminate at dislocation points or exit the domain,
cf. Fig. \ref{Fig:SlipLines}.  

The plastic deformation of any simply connected defect-free subset $\omega$ therefore consists on a relative translation between the material above and below the jump set and is characterized by a vector (Burgers vector) tangent to the slip line. The displacement jump away of the dislocations thus satisfies\footnote{Both statements are equivalent when $\mathbf{X}, \mathbf{X}-\mathbf{b},\mathbf{X}+\mathbf{b} \in \mathcal{J}$.} 
\begin{equation} \label{Eq:Slip}
\begin{split}
&\boldsymbol \varphi^-(\mathbf{X}) =  \boldsymbol \varphi^+(\mathbf{X}-\mathbf{b}), \quad \forall\ \mathbf{X}, \mathbf{X}-\mathbf{b} \in \mathcal{J}, \\
&\boldsymbol \varphi^+(\mathbf{X})= \boldsymbol \varphi^-(\mathbf{X}+ \mathbf{b}), \quad \forall\ \mathbf{X}, \mathbf{X}+\mathbf{b} \in \mathcal{J}.
\end{split}
\end{equation}

\begin{figure}
\begin{center}
    {\includegraphics[width=0.8\textwidth]{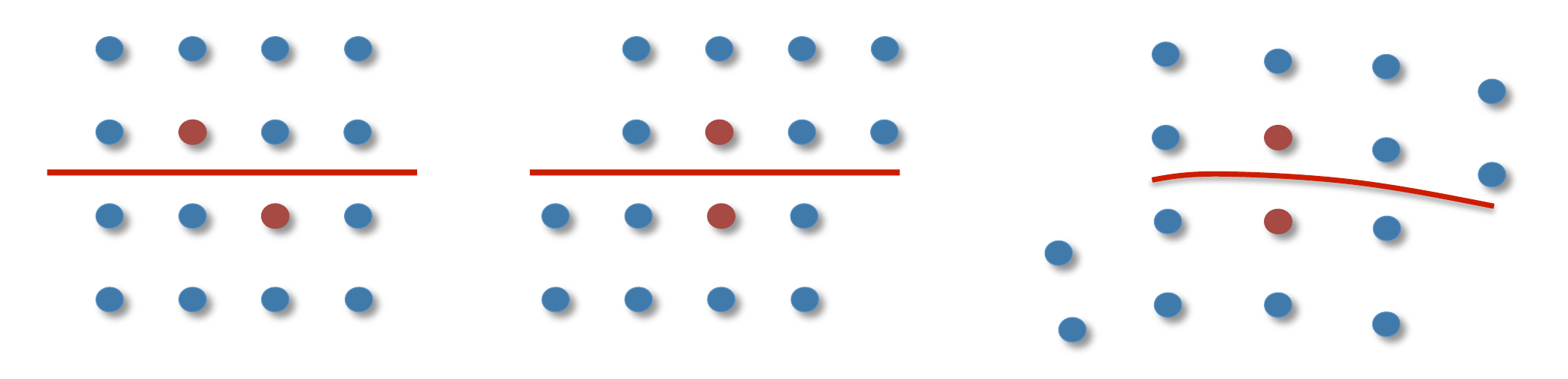}}
    \caption[]{Atomic positions of a perfect crystal (left image) after a  plastic deformation (middle image) and an elastoplastic deformation (right image).}
    \label{Fig:SlipCondition}
\end{center}
\end{figure}

The physical interpretation of this restriction becomes clear when analyzing the position of the atoms close to the jump set prior and after an elastoplastic deformation of a crystalline structure as depicted in Fig. \ref{Fig:SlipCondition}. The Burgers vector, which, in general, corresponds to the interatomic distance in the direction of slip, and the slip line, clearly characterize the plastic deformation with quantities independent of the elastic distortion.

\begin{figure} 
\begin{center}
    {\includegraphics[width=0.75\textwidth]{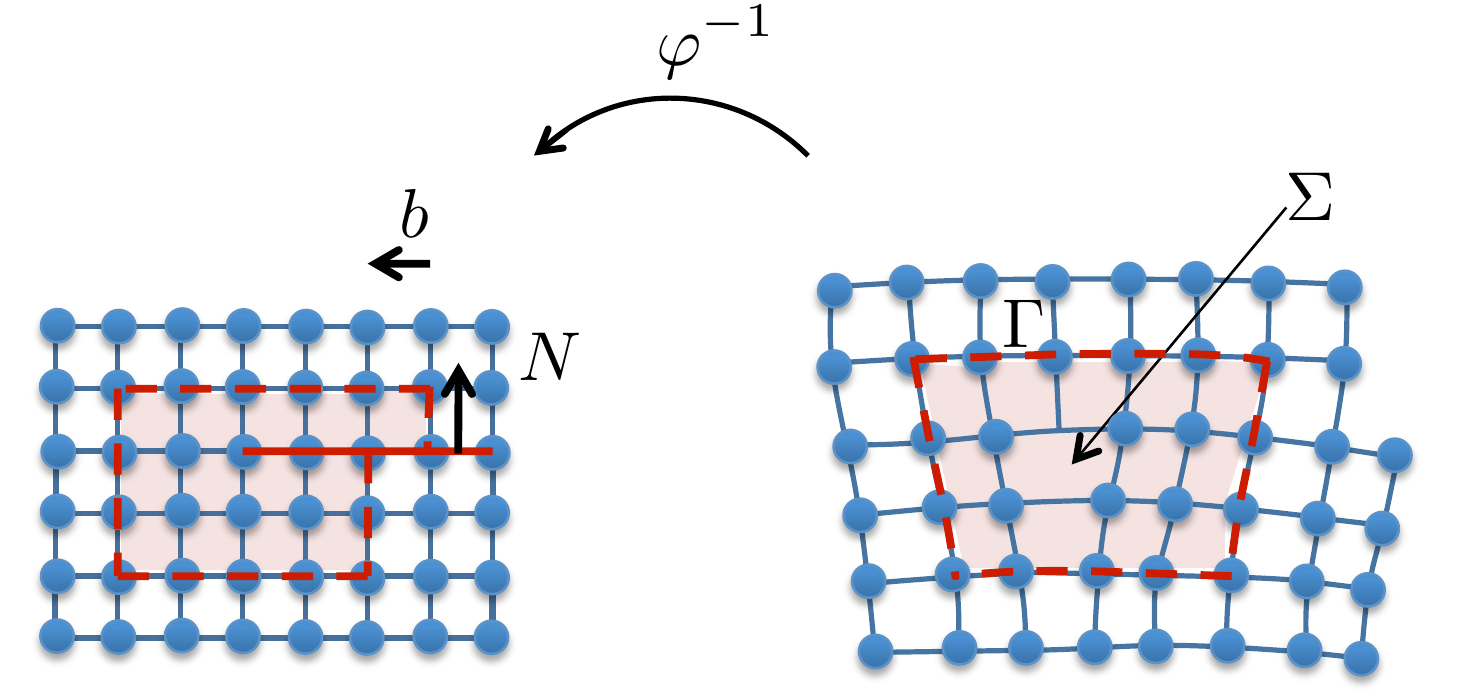}}
    \caption[]{Burgers circuit on the deformed configuration (right image) pulled back to the reference configuration (left image).}
    \label{Fig:BurgersCircuit}
\end{center}
\end{figure}

Equations (\ref{Eq:Slip}) provide an implicit relation between the Burgers
vector and the deformation mapping. However, an approximate explicit relation
can be obtained given sufficient regularity of $\varphi$ and $\mathbf{F}^e$ outside of the jumpset. Indeed, a Taylor expansion gives,
\begin{equation} \label{Eq:phi_jump}
\begin{split}
\llbracket \boldsymbol\varphi \rrbracket (\mathbf{X}) &= \boldsymbol\varphi^+(\mathbf{X}) - \boldsymbol\varphi^+(\mathbf{X}-\mathbf{b}) =\boldsymbol\varphi^+(\mathbf{X}) - \left[ \boldsymbol\varphi^+(\mathbf{X}) - \nabla \boldsymbol\varphi^+(\mathbf{X}) \mathbf{b} - \mathcal{O}(|\mathbf{b}|^2)  \right] \\
& =\boldsymbol\varphi^-(\mathbf{X}+\mathbf{b}) - \boldsymbol\varphi^-(\mathbf{X}) = \left[ \boldsymbol\varphi^-(\mathbf{X})+ \nabla \boldsymbol\varphi^- (\mathbf{X}) \mathbf{b} + \mathcal{O}(|\mathbf{b}|^2)  \right]  - \boldsymbol\varphi^-(\mathbf{X})  \\
&= \frac{1}{2}\Big [\mathbf{F}^{e+} (\mathbf{X}) + \mathbf{F}^{e-}(\mathbf{X}) \Big] \mathbf{b} + \mathcal{O}(|\mathbf{b}|^2) \\
& = \bar{\mathbf{F}}^e(\mathbf{X}) \mathbf{b} + \mathcal{O}(|\mathbf{b}|^2),
\end{split}
\end{equation}
where $\bar{\mathbf{F}}^e(\mathbf{X})$ is defined at $\mathbf{X} \in \mathcal{J}$ as the average of $\mathbf{F}^{e+} (\mathbf{X})$ and $ \mathbf{F}^{e-}(\mathbf{X})$. In the remainder of the paper we will take as a precise representative of $\mathbf{F}^e(\mathbf{X}) $ on $\mathbf{X} \in \mathcal{J}$  the one given by $\mathbf{F}^e(\mathbf{X}) = \bar{\mathbf{F}}^e(\mathbf{X}).$\footnote{$\mathbf{F}^e$ may be arbitrarily defined on sets of zero $\mathcal{L}^2$ measure.}

The kinematic condition of slip can also be expressed as a function of the
inverse mapping via the so-called Burgers circuit\footnote{The deformation
  mappings here considered are invertible on $\Omega \setminus
  \mathcal{J}$.}. To that end, consider a closed circuit in the deformed
configuration that encloses a single dislocation and crosses a
single slip segment. The pullback of the circuit to the reference
configuration, cf.~Fig.~\ref{Fig:BurgersCircuit}, is open and the vector joining its final and
initial point provides the Burgers vector associated to the slip segment,
which is independent of the specific loop chosen as long as the dislocation
content in its interior remains invariant. Mathematically, this can be
expressed with the inverse deformation mapping $\boldsymbol \varphi^{-1}$ and
its corresponding elastic deformation tensor $\mathbf{F}^{e-1}$. More
specifically, the following integral 
\begin{equation}
\int_{\Gamma} \mathbf{F}^{e-1}\cdot d\mathbf{l} = \int_{\Sigma} \text{curl }\mathbf{F}^{e-1}
\end{equation}
is equal to the Burgers vector $\mathbf{b}$ (expressed in the undeformed configuration) if the loop encloses the dislocation and zero otherwise. Therefore,
\begin{equation} \label{Eq:Slip2}
\text{curl } \mathbf{F}^{e-1} = \mathbf{b}\ \delta_{\mathbf{x}_0},
\end{equation}
where $\mathbf{x}_0$ is the position of the dislocation or end of the jump set in the deformed configuration. Equations (\ref{Eq:Slip}) or (\ref{Eq:Slip2}) combined with the orthogonality condition $\mathbf{b}\cdot\mathbf{N}=0$, characterize kinematically the slip condition over non-intersecting jump sets. 

When slip occurs over several intersecting slip systems, the resulting deformation is dependent on the history of dislocation motion throughout the domain. Its kinematic characterization is very similar to that of the deformation induced by a single slip system. On the one hand, close to each dislocation center, an incompatibility condition of the form given by Eq. (\ref{Eq:Slip2}) is satisfied, where $\mathbf{b}$ is the Burgers vector associated to the dislocation, or equivalently, the sum of the Burgers vectors (with appropriate sign) of the slip lines terminating at that point. On the other hand, for any point away from the dislocation points, one can find a neighborhood $\omega$ in which the following relations hold
\begin{equation} \label{Eq:KinematicCondition}
\begin{split}
&\boldsymbol \varphi= \boldsymbol \varphi^e \circ \boldsymbol \varphi^p \\
&\boldsymbol \varphi^e \quad \text{continuous and one-to-one on } \boldsymbol \varphi^p(\omega) \\
& \boldsymbol \varphi^p = \boldsymbol \varphi^{p}_N \circ \boldsymbol \varphi^{p}_{(N-1)} \circ ... \circ \boldsymbol \varphi^{p}_j \circ ... \circ \boldsymbol \varphi^{p}_2 \circ \boldsymbol \varphi^{p}_1 \\
&D \boldsymbol \varphi^{p}_j = \mathbf{I}\ \mathcal{L}^2 + \mathbf{b}_j \otimes \mathbf{N}_j\ \mathcal{H}^1 \lfloor_{\mathcal{J}_j}, \quad \text{on }  \boldsymbol \varphi^{p}_{(j-1)} \circ ... \circ \boldsymbol \varphi^{p}_1 (\omega),
\end{split}
\end{equation}
with $ \mathbf{b}_j \cdot \mathbf{N}_j = 0$ on $\boldsymbol
  \varphi^{p}_{(j-1)} \circ ... \circ \boldsymbol \varphi^{p}_1
  (\omega)$\footnote{In general, the orthogonality condition
      between $ \mathbf{b}_j$ and $\mathbf{N}_j$ will not be satisfied in the
      reference configuration. This will become obvious in the coming
      sections.}. The different intermediate configurations introduced for
the kinematic description ($\boldsymbol \varphi^{p}_{(j-1)} \circ ... \circ
\boldsymbol \varphi^{p}_1 (\omega)$ and $\boldsymbol \varphi^p (\omega)$) are
all realizable and characterize the order in time in which the
  dislocations from different slip systems traversed the given
  subdomain. However, these functions are only local, and global functions
$\boldsymbol \varphi^e$ or $\boldsymbol \varphi^p$ do not exist in
general. The physical interpretation of the above equations, will become clear
in Sections \ref{Sec:Case3} and \ref{Sec:Case4} , where the deformation
induced by two intersecting slip systems is analyzed in detail.
  A single slip system may be activated in several of these steps, as
  discussed in a specific example in Section \ref{sec:Case5} below.

\subsection{Remarks on compatibility and non-interpenetration of matter} \label{Sec:SemicontinousRemarks}

We would like to point out that the notions of compatibility and
  non-interpenetration of matter that are usually considered at the continuum
  scale are not valid, in general, at the microscale. In the continuum
  mechanics framework, a given deformation tensor $\mathbf{F}$ is said to be
  compatible when it derives from a deformation mapping which is continuous
  and one-to-one. This is equivalent for simply connected domains to the
  condition $\text{Curl}\ \mathbf{F} = 0$. However, in the present discrete
  formulation of elastoplasticity, slip is allowed for, and the injectivity
  condition is clearly violated at the jump sets\footnote{Material points that
    belonged to the interior of the domain may also become part of the
    boundary}. Away from the jump set though, the deformation of the body is still described with a continuous deformation mapping, and the standard one-to-one condition applies. In the semicontinuous setting here considered, we then say that a deformation gradient is compatible when it derives from a SBV mapping, which is one-to-one outside of the jumpset and satisfies the kinematic conditions expressed in Eq.~(\ref{Eq:KinematicCondition}).


\section{Examples of elastoplastic deformations in two dimensions} \label{Sec:Examples}

In the previous section, unique definitions for the total deformation gradient $\mathbf{F}$ and the elastic deformation tensor $\mathbf{F}^e$ were provided by means of physical arguments, cf. Eqs. (\ref{Eq:Definition_F}) and (\ref{Eq:Definition_Fe}). Next, we examine several illustrative deformations with increasing complexity in order to infer an appropriate definition for the plastic deformation tensor $\mathbf{F}^p$ within the semicontinuous model. In particular, we consider

\begin{enumerate}
\item[-] a compatible plastic deformation involving a single slip system,
\item[-] a compatible elastoplastic deformation involving a single slip system, 
\item[-] a compatible plastic deformation involving two slip systems,
\item[-] a compatible elastoplastic deformation involving two slip systems,
\item[-] a compatible plastic deformation involving multiple slip systems,
\end{enumerate}
and we also discuss the case of general elastoplastic deformations. For each example we additionally obtain their continuum limit via a sequence of deformations in which the number of slip segments increases to infinity while the lattice parameter ($\epsilon$) tends to zero over a fixed domain, cf. Fig. \ref{Fig:Case1b}. This mathematical limit is physically equivalent to a 'zoom out'  during an experimental observation. The specific scaling used in this process is 
\begin{equation}
|\mathcal{J}|\ \epsilon < C,
\end{equation}
where $|\mathcal{J}|$ measures the length of the jump set in the reference configuration. This scaling ensures that the displacements are at most finite, as shown in Fig. \ref{Fig:Case1b}. Further conditions on the sequence are required in order to guarantee the continuity of the limiting solution as well as to prevent unphysical dislocation structures. However, these technical details are obviated here for the sake of simplicity.

From a notation standpoint, we will denote by $\boldsymbol \varphi_{\epsilon}$ and $\mathcal{J}_{\epsilon}$ the deformation mapping and its jumpset for each value of $\epsilon$ along the sequence, and use $\mathbf{F}_{\epsilon}, \mathbf{F}^e_{\epsilon}, \mathbf{F}^p_{\epsilon}$ to describe the associated discrete deformation tensors. Additionally, the Burgers vector, which is proportional to the lattice parameter, will be denoted by $\mathbf{b}_{\epsilon}$ ($\mathbf{b}_{\epsilon}=\mathbf{b} \epsilon$). The continuum counterparts of the different quantities will be denoted by $\boldsymbol \varphi,\ \mathbf{F},\ \mathbf{F}^e$ and $\mathbf{F}^p$ respectively.

\subsection{Compatible plastic deformation involving a single slip system}
\begin{figure}
\begin{center}
    {\includegraphics[width=0.55\textwidth]{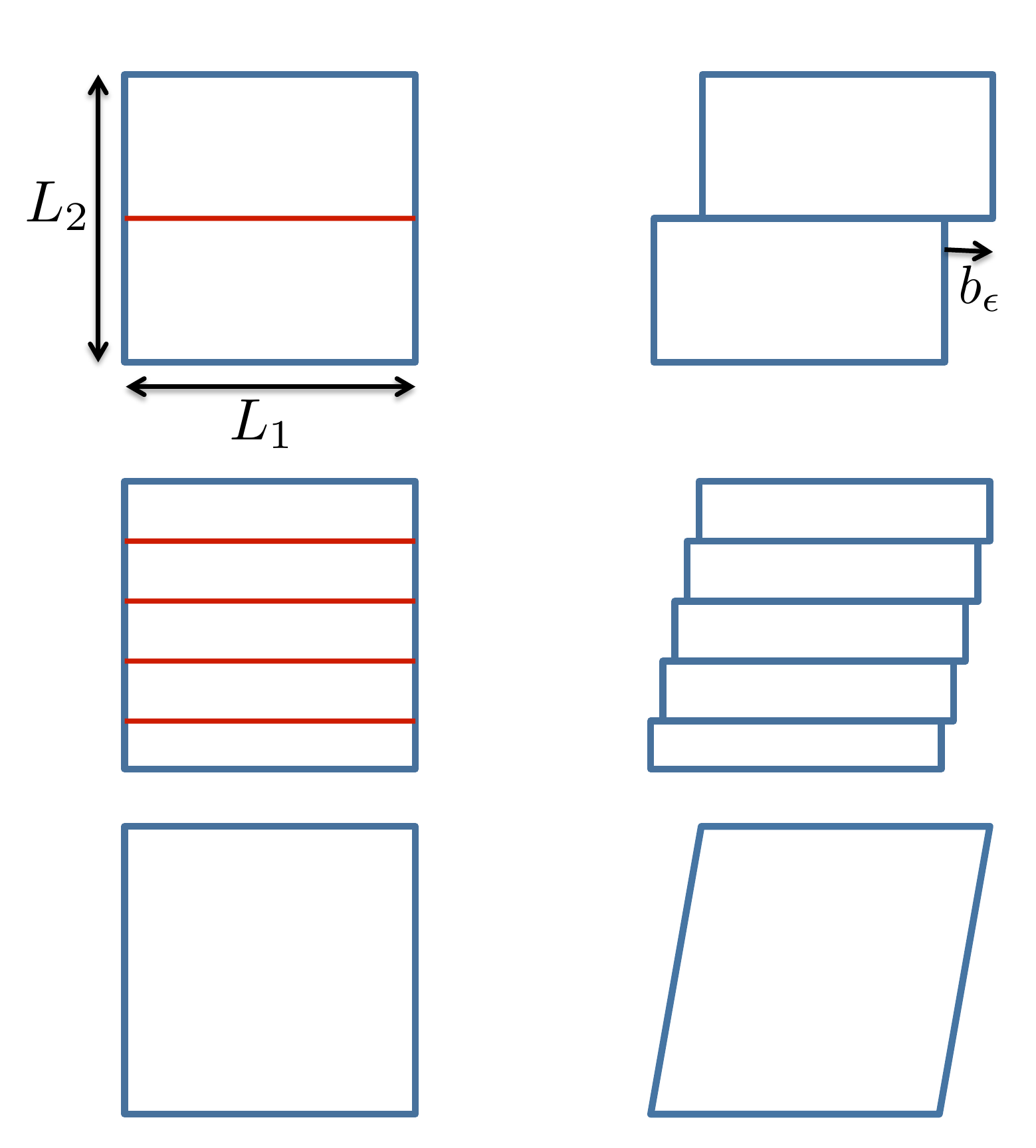}}
    \caption[]{Sequence of deformations converging to a continuous simple shear deformation with a scaling $\frac{N_{\mathcal{J}_{\epsilon}} |\mathbf{b}| \epsilon}{L_2} = \gamma_1$, where $N_{\mathcal{J}_{\epsilon}}$ is the number of slip lines. The reference configurations with the jump sets indicated in red are shown on the left, whereas the deformed configurations are shown on the right.}
    \label{Fig:Case1b}
\end{center}
\end{figure}

We begin by  considering a standard example, namely, an element
of the 
sequence shown in 
Fig.~\ref{Fig:Case1b}. These deformations
which are in the literature sometimes visualized as a deck of
rigid cards, in which many of them are displaced with respect to the
neighbouring one  in the direction $e_1$ by a quantity proportional to the
Burgers vector, 
 clearly do not involve any elastic
distortion. As a result, the plastic deformation tensor is exactly equivalent
to the total deformation gradient, which is  described by 
\begin{equation} \label{Eq:Fp_Case1}
\mathbf{F}^p_{\epsilon} = \mathbf{F}_{\epsilon} = \mathbf{I}\ \mathcal{L}^2 + |\mathbf{b}_{\epsilon}|\ \mathbf{e}_1 \otimes \mathbf{e}_2\ \mathcal{H}^1 \lfloor_{\mathcal{J}_{\epsilon}}
\end{equation}
where  $\mathbf{b}_{\epsilon}=\epsilon |\mathbf{b}| \mathbf{e}_1$.
This sequence converges to a uniform continuous plastic deformation as the number of slip lines $N_{\mathcal{J}}$ increases to infinity and the lattice parameter $\epsilon$ tends to zero, under the scaling $\frac{N_{\mathcal{J}_{\epsilon}} |\mathbf{b}| \epsilon}{L_2} = \gamma_1$. Due to the uniform character of the deformation in the limit, we can obtain such limiting value via volume integration
\begin{equation}
\begin{split}
&\mathbf{F}^p L_1 L_2 = \mathbf{I}\ L_1 L_2 + N_{\mathcal{J}_{\epsilon}} L_1 |\mathbf{b}|\ \epsilon\ \mathbf{e}_1 \otimes \mathbf{e}_2 \\
&\mathbf{F}^p = \mathbf{I} + \gamma_1\ \mathbf{e}_1 \otimes \mathbf{e}_2. \\
\end{split}
\end{equation}

The continuum limit thus corresponds to the expected value for a simple shear deformation. Mathematically, $\mathbf{F}^p_{\epsilon}$ is said to weakly* converge to $\mathbf{F}^p$ ($\mathbf{F}^p_{\epsilon} \overset{*}{\rightharpoonup} \mathbf{F}^p$).

\subsection{Compatible elastoplastic deformation involving a single slip system} \label{Sec:Case2}
\begin{figure} 
\begin{center}
    {\includegraphics[width=0.5\textwidth]{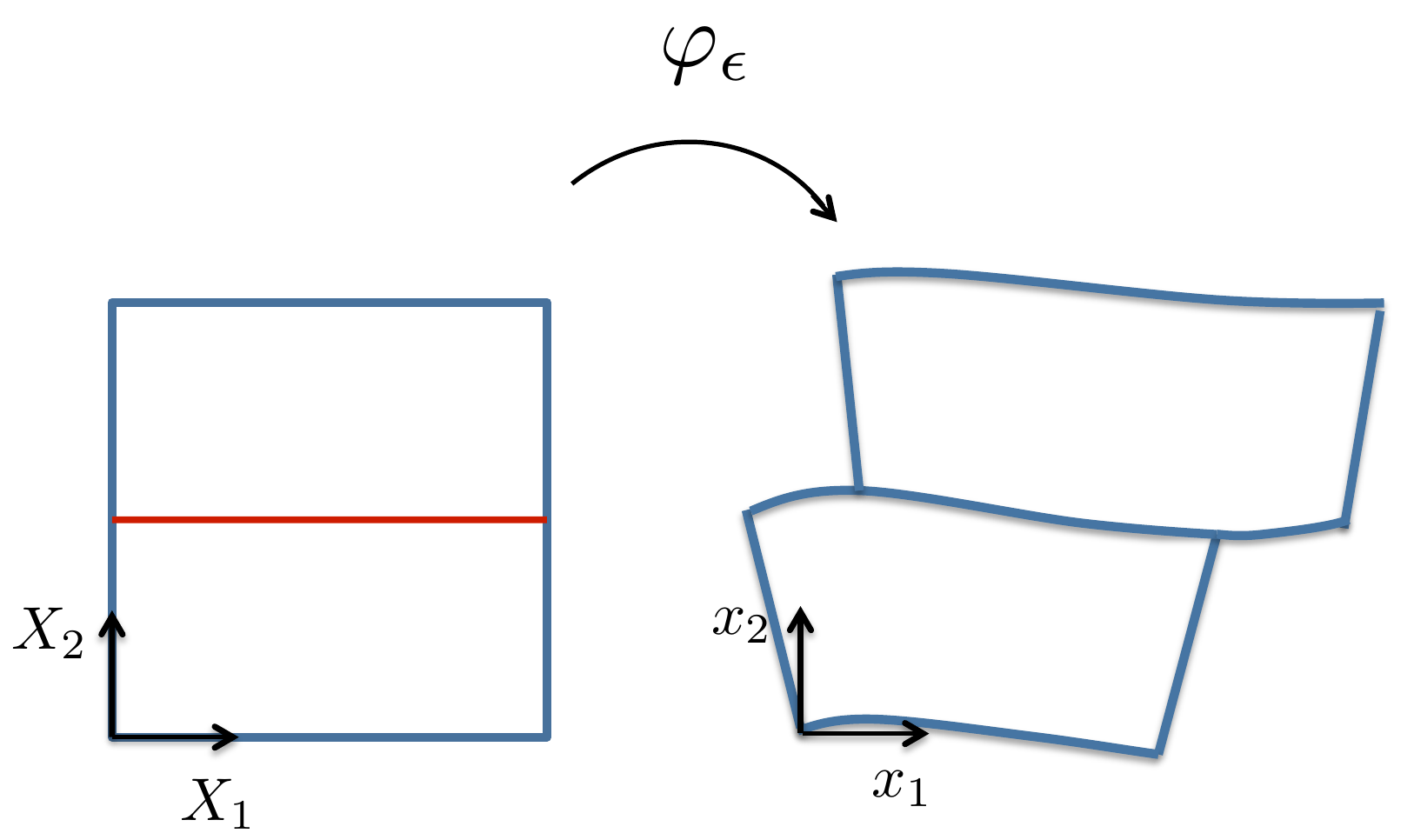}}
    \caption[]{Elastoplastic deformation (right image) induced by slip over the jump set shown in red on the reference configuration (left image).}
    \label{Fig:Case2}
\end{center}
\end{figure}

Next, we consider a sequence of plastic distortions identical to the one in the previous section with a superposed elastic deformation, cf. Fig. \ref{Fig:Case2}. By construction, the measure of shape change induced by slip must equate Eq. (\ref{Eq:Fp_Case1}) and is determined by
\begin{equation} \label{Eq:Case2_Fp}
\mathbf{F}^p_{\epsilon} = \mathbf{I}\ \mathcal{L}^2 + |\mathbf{b}| \epsilon\ \mathbf{e}_1 \otimes \mathbf{e}_2\ \mathcal{H}^1\lfloor_{\mathcal{J}_{\epsilon}},
\end{equation}
which depends exclusively on quantities that are measured in the undeformed configuration. On the other hand, by Eqs. (\ref{Eq:Definition_F}), (\ref{Eq:Definition_Fe}) and (\ref{Eq:phi_jump}), the total deformation gradient satisfies
\begin{equation} \label{Eq:Case2_F}
\mathbf{F}_{\epsilon} = \mathbf{F}^e_{\epsilon} + \llbracket \boldsymbol \varphi_{\epsilon} \rrbracket \otimes \mathbf{e}_2\ \mathcal{H}^1 \lfloor_{\mathcal{J}_{\epsilon}}
\end{equation}
with
\begin{equation}
\llbracket \boldsymbol \varphi_{\epsilon} \rrbracket (\mathbf{X}) =
\mathbf{F}^e_{\epsilon} (\mathbf{X}) \mathbf{b}_{\epsilon} + \mathcal{O}(\mathbf{b}_{\epsilon}^2),
\end{equation}
which combined with Eq. (\ref{Eq:Case2_Fp}), delivers the well-known multiplicative decomposition at the microscale up to an error of order $\epsilon^2$
\begin{equation}
\mathbf{F}_{\epsilon} = \mathbf{F}^e_{\epsilon}\left(\mathbf{I}\ \mathcal{L}^2+ \mathbf{b}_{\epsilon} \otimes \mathbf{N}\ \mathcal{H}^1\lfloor_J \right)+ \mathcal{O}(\mathbf{b}_{\epsilon}^2) = \mathbf{F}^e_{\epsilon} \mathbf{F}^p_{\epsilon} +\mathcal{O}(|\epsilon|^2).
\end{equation}

In this compatible setting, the elastic deformation can be taken constant along the sequence ($\mathbf{F}^e_{\epsilon} = \mathbf{F}^e$), and the limiting deformation gradient then satisfies
\begin{equation}
\mathbf{F} = \mathbf{F}^e \mathbf{F}^p,
\end{equation}
where $\mathbf{F} $ and $\mathbf{F}^p$ are the limits of
$\mathbf{F}_{\epsilon}$ and $\mathbf{F}^p_{\epsilon}$, respectively as
$\epsilon \rightarrow 0$.

\subsection{Compatible plastic deformation involving two slip systems}  \label{Sec:Case3}
\begin{figure}
\begin{center}
    {\includegraphics[width=0.8\textwidth]{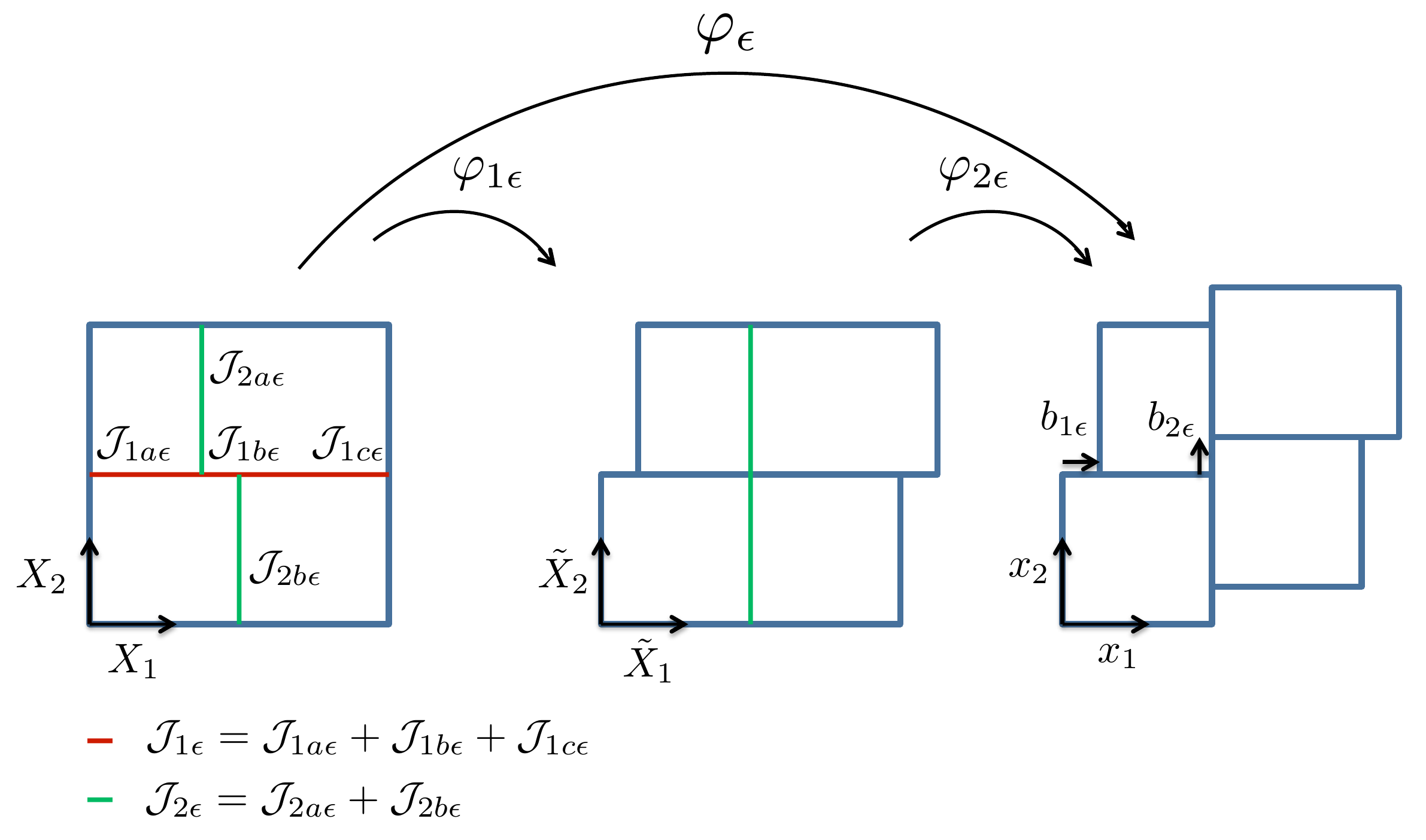}}
    \caption[]{Plastic deformation induced by two successive slips over orthogonal slip systems. The jump set associated to each deformation is shown in the reference configuration (left image).}
    \label{Fig:Case3a}
\end{center}
\end{figure}

Next, we consider a plastic deformation that involves two orthogonal slip systems as shown in Fig. \ref{Fig:Case3a}. The displacement jump over the different segments of the jump set in the reference configuration is
\[ \llbracket \boldsymbol \varphi_{\epsilon} \rrbracket = \llbracket \boldsymbol \varphi_{2\epsilon} \circ \boldsymbol \varphi_{1\epsilon} \rrbracket=\left \{ \begin{array} {l l }
\mathbf{b}_{1\epsilon} & \text{on } J_{1a\epsilon} \cup J_{1c\epsilon} \\
\mathbf{b}_{2\epsilon} & \text{on } J_{2a\epsilon} \cup J_{2b\epsilon} \\
\mathbf{b}_{1\epsilon} + \mathbf{b}_{2\epsilon} & \text{on } J_{1b\epsilon}.
\end{array} \right.
\]

Note that the vertical slip line associated to $\boldsymbol \varphi_{2\epsilon}$ has been pulled back according to $\boldsymbol \varphi_{1\epsilon}$ in order to have a full characterization of the jump set in the reference configuration. This induces a discontinuity in the vertical line of magnitude equal to $\mathbf{b}_{1\epsilon}$. We will refer later on to these small segments as kinks. Due to the  absence of any elastic distortion, the deformation gradient is equivalent to the plastic deformation tensor, and is uniquely defined from the microstructure as
\begin{equation} \label{Eq:Case3_Fp}
\mathbf{F}^p_{\epsilon} = \mathbf{I}\ \mathcal{L}^2 + |\mathbf{b}_{1\epsilon} |\ \mathbf{e}_1\otimes \mathbf{e}_2\ \mathcal{H}^1\lfloor_{\mathcal{J}_{1\epsilon}} + |\mathbf{b}_{2\epsilon} |\ \mathbf{e}_2\otimes \mathbf{e}_1\ \mathcal{H}^1\lfloor_{\mathcal{J}_{2\epsilon}} +  |\mathbf{b}_{2\epsilon} | \mathbf{e}_2\otimes \mathbf{e}_2\ \mathcal{H}^1\lfloor_{\mathcal{J}_{1b\epsilon}} ,
\end{equation}
where $\mathcal{J}_{1\epsilon} = \mathcal{J}_{1a\epsilon} \cup \mathcal{J}_{1b\epsilon} \cup \mathcal{J}_{1c\epsilon}$ and $\mathcal{J}_{2\epsilon} = \mathcal{J}_{2a\epsilon} \cup \mathcal{J}_{2b\epsilon}$.

\begin{figure}
\begin{center}
    {\includegraphics[width=0.8\textwidth]{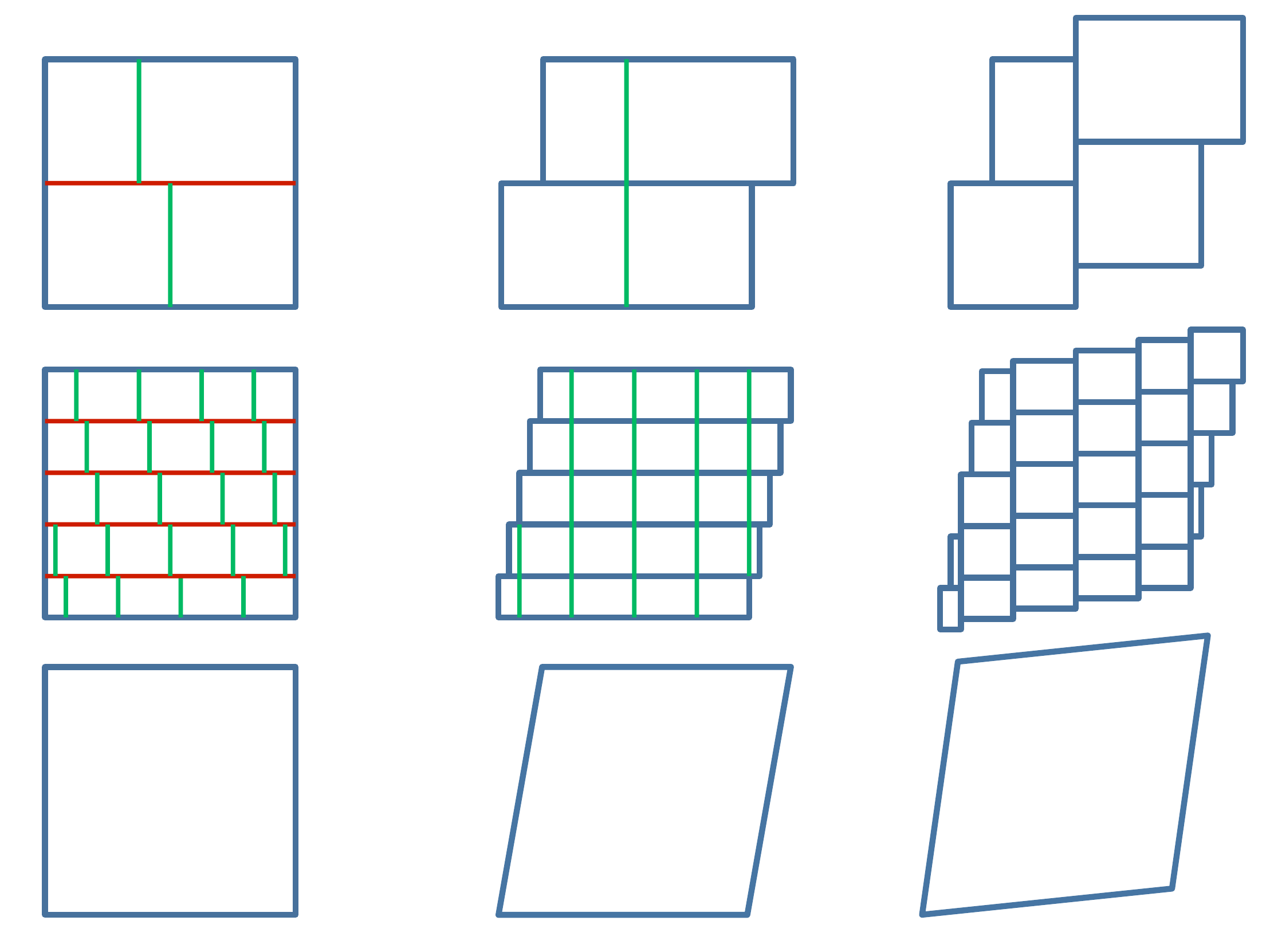}}
    \caption[]{Sequence of plastic deformations involving two orthogonal slip systems satisfying the scalings $\gamma_1 = \frac{N_{\mathcal{J}_{1\epsilon}} |\mathbf{b}_{1\epsilon}| }{L_2}$ and $\gamma_2 = \frac{N_{\mathcal{J}_{2\epsilon}} |\mathbf{b}_{2\epsilon}| }{L_1}$. $N_{\mathcal{J}_{1\epsilon}}$ and $N_{\mathcal{J}_{2\epsilon}}$ denote the number of slip lines in each direction, and $\mathbf{b}_{1\epsilon}$ and $\mathbf{b}_{2\epsilon}$, the slip on each set of jump sets. The left column of images indicates the reference configuration where the jump sets are highlighted. The deformations resulting from slip on the first system is shown in the middle column, and the final configuration is shown on the right column. }
    \label{Fig:Case3b}
\end{center}
\end{figure}

We again consider a sequence of such deformations converging to a continuous distortion, c.f. Fig. \ref{Fig:Case3b}, satisfying the following scalings
\begin{equation}
\begin{split}
&\gamma_1 = \frac{N_{\mathcal{J}_{1\epsilon}} |\mathbf{b}_{1\epsilon}| }{L_2}\\
&\gamma_2 = \frac{N_{\mathcal{J}_{2\epsilon}} |\mathbf{b}_{2\epsilon}| }{L_1}, \\
\end{split}
\end{equation}
where $N_{\mathcal{J}_{1\epsilon}}$ and $N_{\mathcal{J}_{2\epsilon}}$ represent the number of slip lines associated to $\boldsymbol \varphi_{1\epsilon}$ and $\boldsymbol \varphi_{2\epsilon}$ respectively. In view of the uniformity of the solution, the limiting plastic deformation can be obtained by volumetric averaging
\begin{equation}
\begin{split}
&\mathbf{F}^p L_1 L_2 = \mathbf{I}\ L_1 L_2 + N_{\mathcal{J}_{1\epsilon}} L_1 |\mathbf{b}_{1\epsilon}| \ \mathbf{e}_1 \otimes \mathbf{e}_2 + N_{\mathcal{J}_{2\epsilon}} L_2 |\mathbf{b}_{2\epsilon}|\ \mathbf{e}_2 \otimes \mathbf{e}_1 \\
&\phantom{\mathbf{F}^p L_1 L_2 =}+ N_{\mathcal{J}_{1\epsilon}} N_{\mathcal{J}_{2\epsilon}} |\mathbf{b}_{1\epsilon}| |\mathbf{b}_{2\epsilon}| \ \mathbf{e}_2 \otimes \mathbf{e}_2 \\
&\mathbf{F}^p = \mathbf{I} + \gamma_1 \mathbf{e}_1 \otimes \mathbf{e}_2+ \gamma_2 \mathbf{e}_2 \otimes \mathbf{e}_1 + \gamma_1\gamma_2 \mathbf{e}_2 \otimes \mathbf{e}_2. 
\end{split}
\end{equation}

This continuum limit corresponds to the expected composition of two (compatible) simple shears at the macroscopic scale: $\mathbf{F}^p_1 = \mathbf{I} + \gamma_1 \mathbf{e}_1 \otimes \mathbf{e}_2$ and $\mathbf{F}^p_2 = \mathbf{I} + \gamma_2 \mathbf{e}_2 \otimes \mathbf{e}_1$. That is,
\begin{equation}
\begin{split}
\mathbf{F}^p &= \mathbf{F}^p_2 \mathbf{F}^p_1 = \left(\mathbf{I} + \gamma_2 \mathbf{e}_2 \otimes \mathbf{e}_1 \right)\left(\mathbf{I} + \gamma_1 \mathbf{e}_1 \otimes \mathbf{e}_2 \right) \\
&= \mathbf{I} + \gamma_1 \mathbf{e}_1 \otimes \mathbf{e}_2+ \gamma_2 \mathbf{e}_2 \otimes \mathbf{e}_1 + \gamma_1\gamma_2 \mathbf{e}_2 \otimes \mathbf{e}_2. \\
\end{split}
\end{equation}

This simple example shows how the kinks associated to the pull-back of the slip lines of $\boldsymbol \varphi_{2\epsilon}$ give rise to the second order term in the total plastic deformation tensor, and are the ones responsible for the non-commutative character of the individual shears. 

\begin{figure}
\begin{center}
    {\includegraphics[width=0.8\textwidth]{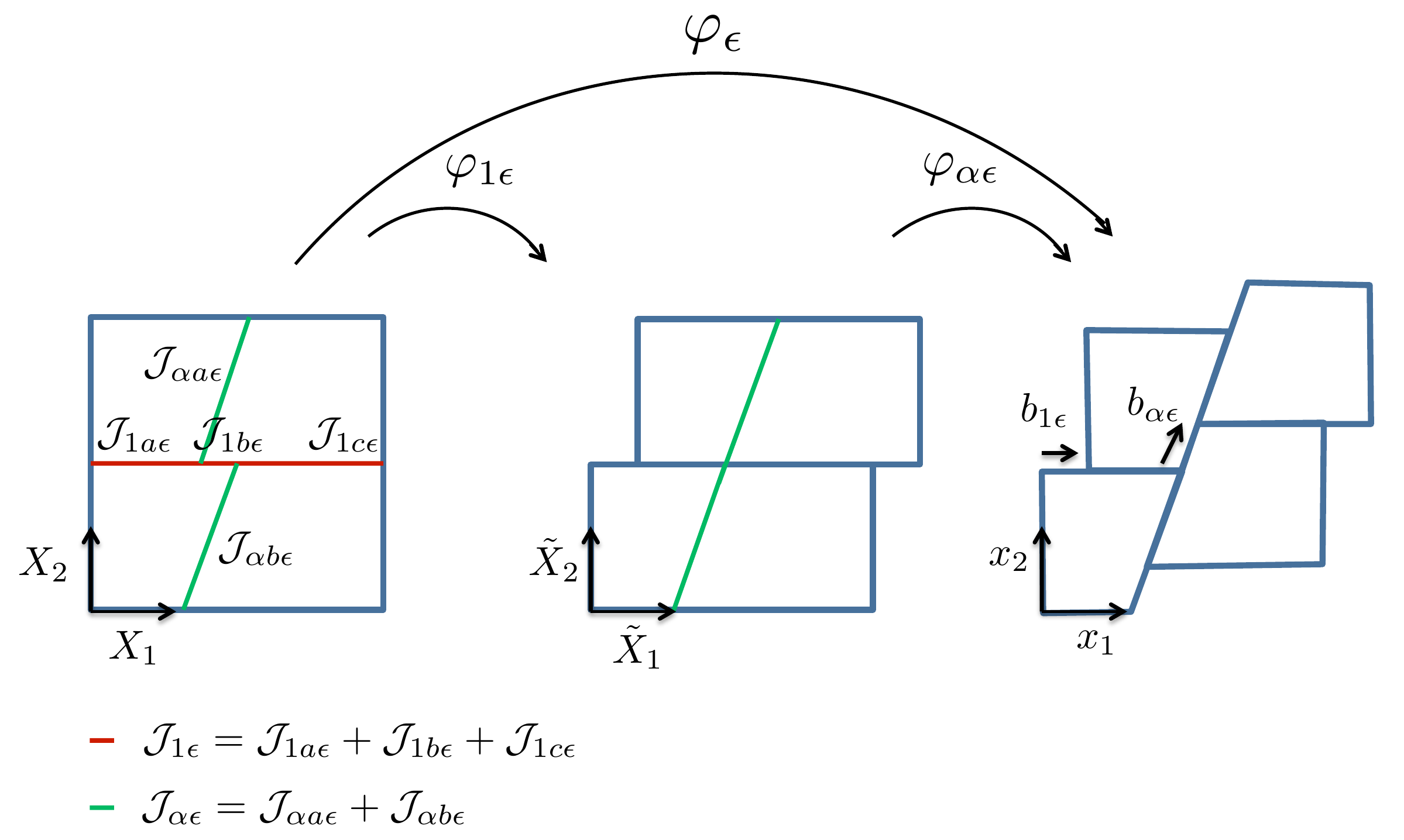}}
    \caption[]{Plastic deformation induced by two successive slips over non-orthogonal slip systems. The jump set associated to each deformation is shown in the reference configuration (left image).}
    \label{Fig:Case3e}
\end{center}
\end{figure}

\begin{figure}
\begin{center}
    {\includegraphics[width=0.8\textwidth]{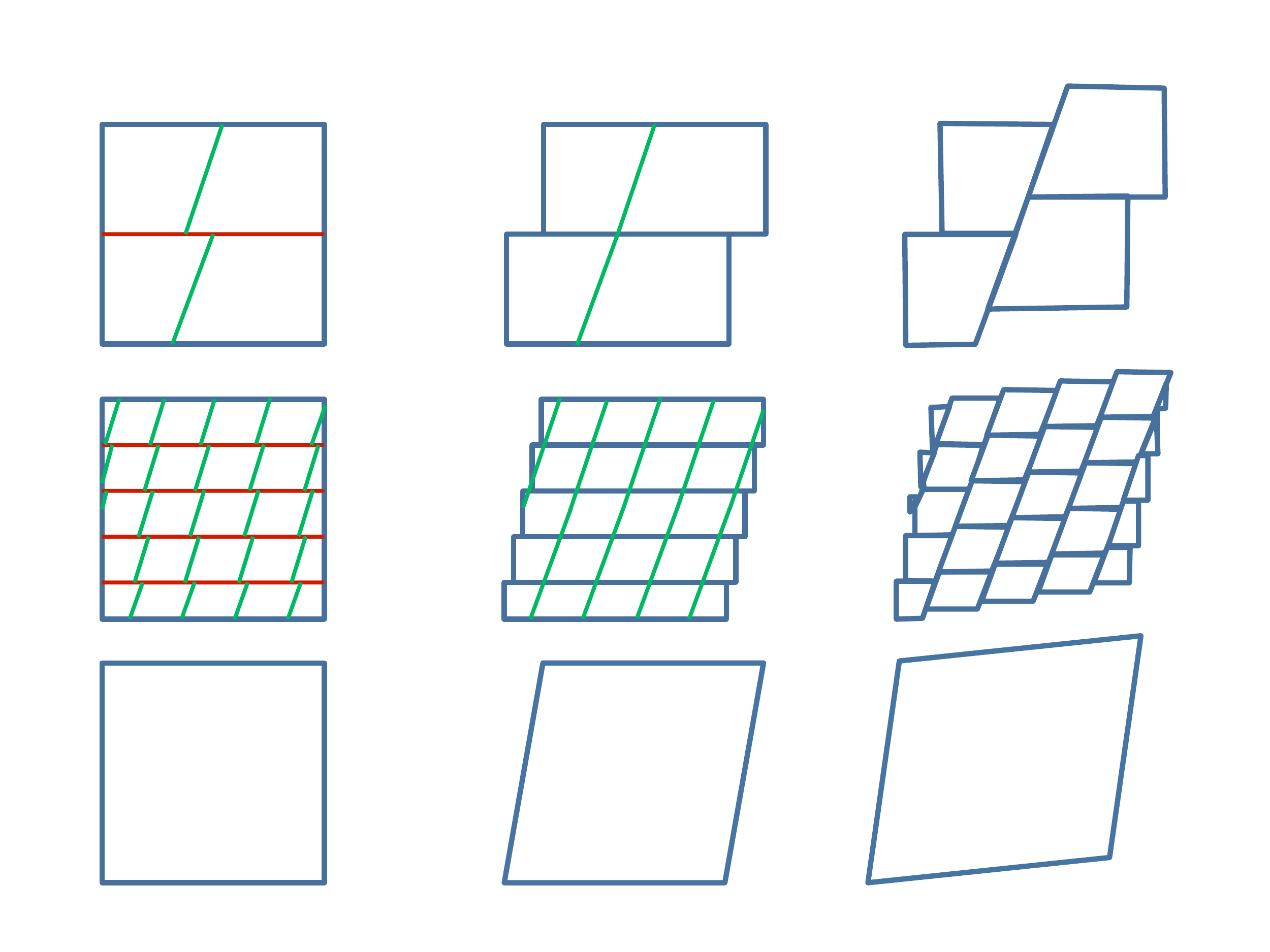}}
    \caption[]{Sequence of plastic deformations involving two non-orthogonal slip systems satisfying the scalings $\gamma_1 = \frac{N_{\mathcal{J}_{1\epsilon}} |\mathbf{b}_{1\epsilon}| }{L_2}$ and $\gamma_{\alpha} = \frac{N_{\mathcal{J}_{\alpha \epsilon}} |\mathbf{b}_{\alpha\epsilon}| }{L_{\alpha}^{\bot}}$. $N_{\mathcal{J}_{1\epsilon}}$ and $N_{\mathcal{J}_{\alpha \epsilon}}$ denote the number of slip lines in each direction, and $\mathbf{b}_{1\epsilon}$ and $\mathbf{b}_{\alpha \epsilon}$, the slip on each set of jump sets. The left column of images indicates the reference configuration where the jump sets are highlighted. The deformations resulting from slip on the first system is shown in the middle column, and the final configuration is shown on the right column. }
    \label{Fig:Case3d}
\end{center}
\end{figure}

The example considered here assumes that the two slip systems are orthogonal for clarity in the exposition. However, the results hold for an arbitrary angle $\alpha$ between the two slip systems. Indeed, for the sequence shown in Fig. \ref{Fig:Case3d}, the displacement jump and discrete plastic deformation tensor read, c.f. Fig. \ref{Fig:Case3e},
\[ \llbracket \boldsymbol \varphi_{\epsilon} \rrbracket = \llbracket \boldsymbol \varphi_{\alpha \epsilon} \circ \boldsymbol \varphi_{1\epsilon} \rrbracket=\left \{ \begin{array} {l l }
\mathbf{b}_{1\epsilon} & \text{on } J_{1a\epsilon} \cup J_{1c\epsilon} \\
\mathbf{b}_{\alpha \epsilon} & \text{on } J_{\alpha a \epsilon} \cup J_{\alpha b\epsilon} \\
\mathbf{b}_{1\epsilon} + \mathbf{b}_{\alpha \epsilon} & \text{on } J_{1b\epsilon}
\end{array} \right.
\]
\begin{equation}
\mathbf{F}^p_{\epsilon} = \mathbf{I}\ \mathcal{L}^2 + |\mathbf{b}_{1\epsilon} |\ \mathbf{e}_1\otimes \mathbf{e}_2\ \mathcal{H}^1\lfloor_{\mathcal{J}_{1\epsilon}} + |\mathbf{b}_{\alpha\epsilon} |\ \mathbf{e}_{\alpha}^{\bot}\otimes \mathbf{e}_{\alpha}\ \mathcal{H}^1\lfloor_{\mathcal{J}_{\alpha \epsilon}} +  |\mathbf{b}_{\alpha\epsilon} | \mathbf{e}_2\otimes \mathbf{e}_{\alpha}^{\bot}\ \mathcal{H}^1\lfloor_{\mathcal{J}_{1b\epsilon}},
\end{equation}
where $\mathbf{e}_{\alpha}$ is a unit vector parallel to $\mathbf{b}_{\alpha \epsilon}$.

We denote by $\beta$ the spacing between slip lines of the second plastic deformation, and define
\begin{equation}
\begin{split}
&L_{\alpha} = \frac{\sum_i^{N_{\mathcal{J}_{\alpha}}} |\mathcal{J}_{\alpha i}|}{ N_{\mathcal{J}_{\alpha \epsilon}} } \\
&L_{\alpha}^{\bot} = \beta N_{\mathcal{J}_{\alpha \epsilon}}.\\
\end{split}
\end{equation}

Then,
\begin{equation}
L_1 L_2 = \beta  \left(\sum_i^{N_{\mathcal{J}_{\alpha}}} |\mathcal{J}_{\alpha i}| \right) + \mathcal{O}( N_{\mathcal{J}_{\alpha \epsilon}}\beta^2) = L_{\alpha} L_{\alpha }^{\bot} + \mathcal{O}(N_{\mathcal{J}_{\alpha \epsilon}}\beta^2).
\end{equation}

The sequence under consideration satisfies the following scaling
\begin{equation}
\begin{split}
&\gamma_1 = \frac{N_{\mathcal{J}_{1\epsilon}} |\mathbf{b}_{1\epsilon}| }{L_2}\\
&\gamma_{\alpha} = \frac{N_{\mathcal{J}_{\alpha \epsilon}} |\mathbf{b}_{\alpha\epsilon}| }{L_{\alpha}^{\bot}}, \\
\end{split}
\end{equation}
and converges to a continuous deformation, which can be obtained via volumetric averaging
\begin{equation}
\begin{split}
&\mathbf{F}^p L_1 L_2 = \mathbf{I}\ L_1 L_2 + N_{\mathcal{J}_{1\epsilon}} L_1 |\mathbf{b}_{1\epsilon}| \ \mathbf{e}_1 \otimes \mathbf{e}_2 +  \left(\sum_i^{ N_{\mathcal{J}_{\alpha \epsilon}}} |\mathcal{J}_{\alpha i}| \right)  |\mathbf{b}_{\alpha\epsilon}|\ \mathbf{e}_{\alpha}^{\bot} \otimes \mathbf{e}_{\alpha} \\
&\phantom{\mathbf{F}^p L_1 L_2 =}+ N_{\mathcal{J}_{1\epsilon}} N_{\mathcal{J}_{\alpha\epsilon}} |\mathbf{b}_{1\epsilon}| |\mathbf{b}_{\alpha\epsilon}| \ \mathbf{e}_{\alpha}^{\bot} \otimes \mathbf{e}_2 \\
&\mathbf{F}^p = \mathbf{I} + \gamma_1 \mathbf{e}_1 \otimes \mathbf{e}_2+ \gamma_{\alpha} \mathbf{e}_{\alpha}^{\bot} \otimes \mathbf{e}_{\alpha} + \gamma_1\gamma_{\alpha} (\mathbf{e}_1 \cdot \mathbf{e}_{\alpha}) \mathbf{e}_{\alpha}^{\bot} \otimes \mathbf{e}_2. 
\end{split}
\end{equation}

The result is, again, the one expected form two continuous successive plastic deformations
\begin{equation}
\begin{split}
\mathbf{F}^p &= \mathbf{F}^p_{\alpha} \mathbf{F}^p_1 = \left(\mathbf{I} + \gamma_{\alpha} \mathbf{e}_{\alpha}^{\bot} \otimes \mathbf{e}_{\alpha} \right)\left(\mathbf{I} + \gamma_1 \mathbf{e}_1 \otimes \mathbf{e}_2 \right) \\
&= \mathbf{I} + \gamma_1 \mathbf{e}_1 \otimes \mathbf{e}_2+ \gamma_{\alpha} \mathbf{e}_{\alpha}^{\bot} \otimes \mathbf{e}_{\alpha} + \gamma_1\gamma_{\alpha} (\mathbf{e}_1 \cdot \mathbf{e}_{\alpha}) \mathbf{e}_{\alpha}^{\bot} \otimes \mathbf{e}_2. \\
\end{split}
\end{equation}

\subsection{Compatible elastoplastic deformation involving two slip systems} \label{Sec:Case4}

\begin{figure}
\begin{center}
    {\includegraphics[width=0.8\textwidth]{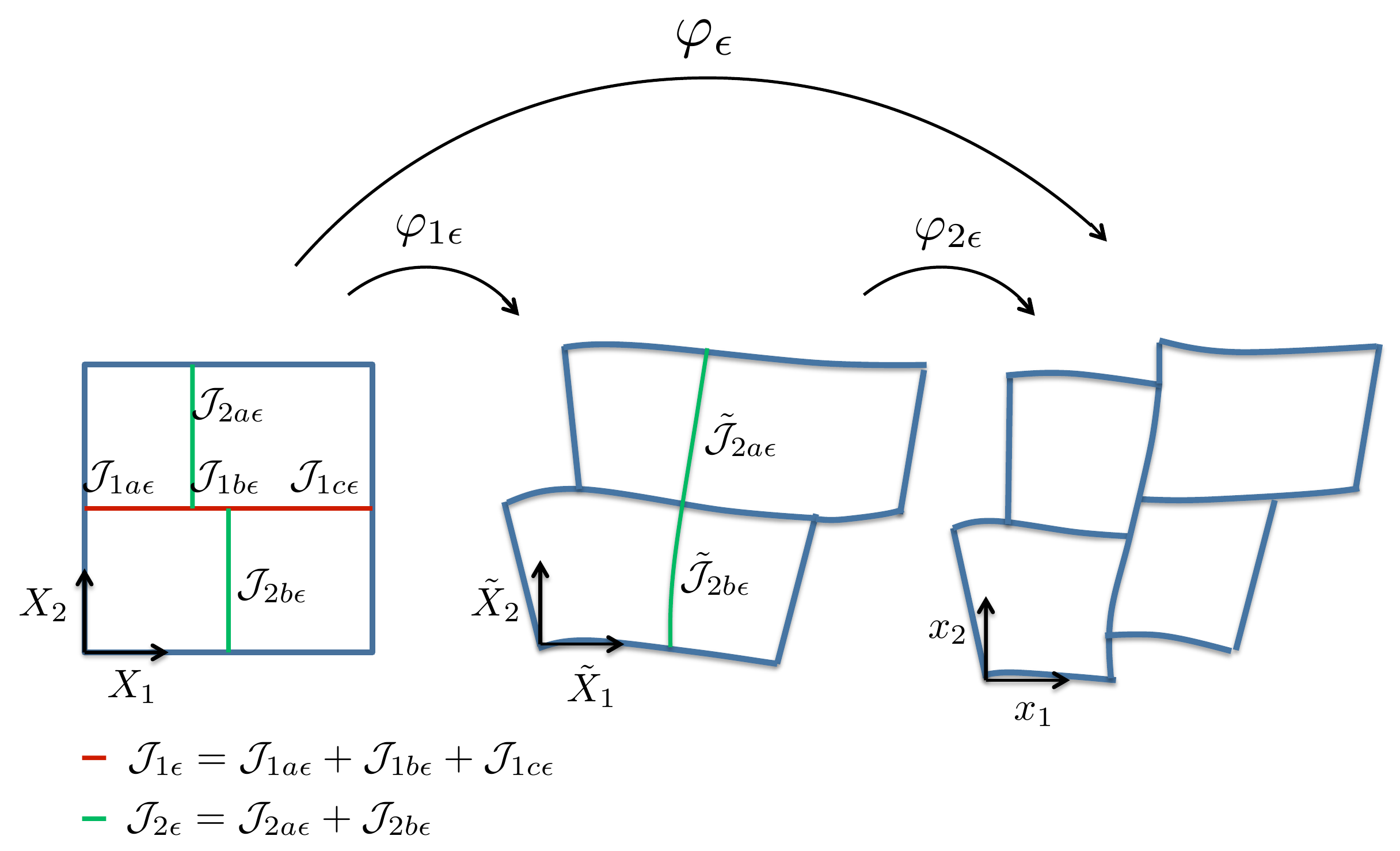}}
    \caption[]{Composition of two elastoplastic deformations. The jump sets associated to each deformation are labeled in the reference configuration (left image).}
    \label{Fig:Case4a}
\end{center}
\end{figure}

Next, we analyze a sequence of deformations induced by two orthogonal slip systems similar to that of Fig.~\ref{Fig:Case3b}, together with a superposed elastic deformation at each step of the plastic distortion, c.f.~Fig.~\ref{Fig:Case4a}. Note that the slip line associated to the $\boldsymbol \varphi_{2\epsilon}$ is in general curved due to the elastic deformation induced by $\boldsymbol \varphi_{1\epsilon}$. However, its pullback to the reference configuration necessarily consists of the union of straight segments due to the kinematic and crystallographic reasons discussed in Sec.~\ref{Sec:SemiContinuousModel}.

$\boldsymbol \varphi_{1\epsilon}$ and $\boldsymbol \varphi_{2\epsilon}$ are both SBV functions, whose gradients can be written as
\begin{equation}
\begin{split}
&\mathbf{F}_{1\epsilon} =D\boldsymbol \varphi_{1\epsilon} = \nabla \boldsymbol \varphi_{1\epsilon}\ \mathcal{L}^2 + \llbracket \boldsymbol \varphi_{1\epsilon} \rrbracket \otimes \mathbf{N}_{1\epsilon}\ \mathcal{H}^1 \lfloor_{\mathcal{J}_{1\epsilon}}, \quad \forall\ \mathbf{X} \in \Omega \\
&\tilde{\mathbf{F}}_{2\epsilon} = \tilde{D}\boldsymbol \varphi_{2\epsilon} = \tilde{\nabla} \boldsymbol \varphi_{2\epsilon}\ \mathcal{L}^2 +  \llbracket \boldsymbol \varphi_{2\epsilon} \rrbracket \otimes \tilde{\mathbf{N}}_{2\epsilon}\ \mathcal{H}^1 \lfloor_{\tilde{\mathcal{J}}_{2\epsilon}}, \quad \forall\ \tilde{\mathbf{X}} \in \boldsymbol \varphi_{1\epsilon} (\Omega), \\
\end{split}
\end{equation}
where the tilde is used to represent quantities in the configuration resulting
from the first mapping and ${\mathbf{N}}_{\epsilon}$ denotes the unit normal to 
$\mathcal{J}_{\epsilon}$. By Eq. (\ref{Eq:phi_jump}), $ \llbracket  \boldsymbol
\varphi_{1\epsilon} \rrbracket $ can be approximated up to second order terms
in $\epsilon$ by 
\begin{equation}
\llbracket \boldsymbol \varphi_{1\epsilon} \rrbracket \simeq \mathbf{F}^e_{1\epsilon} \mathbf{b}_{1\epsilon}, \quad \text{on } \mathcal{J}_{1\epsilon}.
\end{equation}

Similarly, by reasoning with $\boldsymbol \varphi_{\epsilon} = \boldsymbol \varphi_{2\epsilon} \circ \boldsymbol \varphi_{1 \epsilon} $ on points in $\mathcal{J}_{2\epsilon}$, one obtains
\begin{equation}
\llbracket \boldsymbol \varphi_{2\epsilon} \circ \boldsymbol \varphi_{1\epsilon} \rrbracket \simeq \mathbf{F}^e_{2\epsilon} \mathbf{F}^e_{1\epsilon} \mathbf{b}_{2\epsilon}, \quad \text{on }\mathcal{J}_{2\epsilon}.
\end{equation}
Finally, the normal $\tilde{\mathbf{N}}_{2\epsilon}$ can also be referred to the reference configuration via the standard change in area induced by the first elastic deformation
\begin{equation}
\tilde{\mathbf{N}}_{2\epsilon}\ \mathcal{H}^1 \lfloor_{\tilde{\mathcal{J}}_{2 \epsilon}} \simeq \det( \mathbf{F}^e_{1\epsilon}) \mathbf{F}^{e-T}_{1\epsilon} \mathbf{N}_{2\epsilon}\ \mathcal{H}^1 \lfloor_{\tilde{\mathcal{J}}_{2\epsilon}} = \mathbf{F}^{e-T}_{1\epsilon} \mathbf{N}_{2\epsilon}\ \mathcal{H}^1 \lfloor_{\mathcal{J}_{2\epsilon}}.
\end{equation}

Combining all these elements, we obtain that the two deformation gradients can be approximated by
\begin{equation}
\begin{split}
\mathbf{F}_{1\epsilon} &\simeq \mathbf{F}^e_{1\epsilon} \left( \mathbf{I}\ \mathcal{L}^2 + \mathbf{b}_{1\epsilon} \otimes \mathbf{N}_{1\epsilon} \mathcal{H}^1\lfloor_{\mathcal{J}_{1\epsilon}} \right) =  \mathbf{F}^e_{1\epsilon}  \mathbf{F}^p_{1\epsilon} \\
\mathbf{F}_{2\epsilon} &= \tilde{\mathbf{F}}_{2\epsilon}(\boldsymbol \varphi_{1\epsilon}(\mathbf{X})) \simeq \mathbf{F}^e_{2\epsilon} \left( \mathbf{I}\ \mathcal{L}^2 + \mathbf{F}^e_{1\epsilon} \mathbf{b}_{2\epsilon} \otimes \mathbf{N}_{2\epsilon}\ \mathbf{F}^{e-1}_{1\epsilon}\ \mathcal{H}^1\lfloor_{\mathcal{J}_{2\epsilon}} \right) \\
& = \mathbf{F}^e_{2\epsilon} \mathbf{F}^e_{1\epsilon} \left( \mathbf{I}\ \mathcal{L}^2 + \mathbf{b}_{2\epsilon} \otimes \mathbf{N}_{2\epsilon}\ \mathcal{H}^1\lfloor_{\mathcal{J}_{2\epsilon}} \right) \mathbf{F}^{e-1}_{1\epsilon} = \mathbf{F}^e_{2\epsilon}  \mathbf{F}^e_{1\epsilon} \mathbf{F}^p_{2\epsilon} \mathbf{F}^{e-1}_{1\epsilon},\\
\end{split}
\end{equation}
and their product satisfies
\begin{equation} \label{Eq:Case4_Fe2Fp2Fe1Fp1}
\mathbf{F}_{2\epsilon} \mathbf{F}_{1\epsilon} \simeq \left( \mathbf{F}^e_{2\epsilon}  \mathbf{F}^e_{1\epsilon} \mathbf{F}^p_{2\epsilon} \mathbf{F}^{e-1}_{1\epsilon} \right) \left( \mathbf{F}^e_{1\epsilon}  \mathbf{F}^p_{1\epsilon}\right) = \mathbf{F}^e_{2\epsilon}  \mathbf{F}^e_{1\epsilon} \mathbf{F}^p_{2\epsilon} \mathbf{F}^p_{1\epsilon}.
\end{equation}

We see that, again, by pulling back all the quantities associated to the plastic slip to the reference configuration, the expected decomposition of the deformation mapping is recovered up to higher order terms, with the total elastic deformation on the left of the expression and the total plastic deformation on the right. The exact  multiplicative decomposition in the continuum limit can be easily shown by maintaining the elastic deformations invariant along the sequence ($\mathbf{F}^e_{1\epsilon} = \mathbf{F}^e_1$ and $\mathbf{F}^e_{2\epsilon} = \mathbf{F}^e_2$)
\begin{equation} \label{Eq:Case4_FeFp}
\mathbf{F} = \mathbf{F}^e\mathbf{F}^p,
\end{equation}
where $ \mathbf{F}$ and $\mathbf{F}^p$ are the limits of $\mathbf{F}_{\epsilon}$ and $\mathbf{F}^p_{\epsilon}$ respectively, with $\mathbf{F}^p_{\epsilon}$ as in Eq. (\ref{Eq:Case3_Fp}).

For completeness, we point out that the product of two singular measures such as $\mathbf{F}_{2\epsilon} \mathbf{F}_{1\epsilon}$ or $\mathbf{F}^p_{2\epsilon} \mathbf{F}^p_{1\epsilon}$ is not well-defined in general. However, the final result given by Eq. (\ref{Eq:Case4_FeFp}) is mathematically sound. It can be derived by realizing that the deformation  $\boldsymbol \varphi_{\epsilon} = \tilde{\boldsymbol \varphi}^e_{2\epsilon} \circ \tilde{\boldsymbol \varphi}^p_{2\epsilon} \circ \boldsymbol \varphi^e_{1\epsilon} \circ \boldsymbol \varphi^p_{1\epsilon}$ is equivalent to a deformation of the form $\boldsymbol \varphi^e_{2\epsilon} \circ \boldsymbol \varphi^e_{1\epsilon} \circ \boldsymbol \varphi^p_{2\epsilon} \circ \boldsymbol \varphi^p_{1\epsilon} $, c.f. kinematic conditions given by Eqs.~(\ref{Eq:KinematicCondition}).

\subsection{Compatible plastic deformation involving multiple slip systems}\label{sec:Case5}
\begin{figure}
\begin{center}
    {\includegraphics[width=0.8\textwidth]{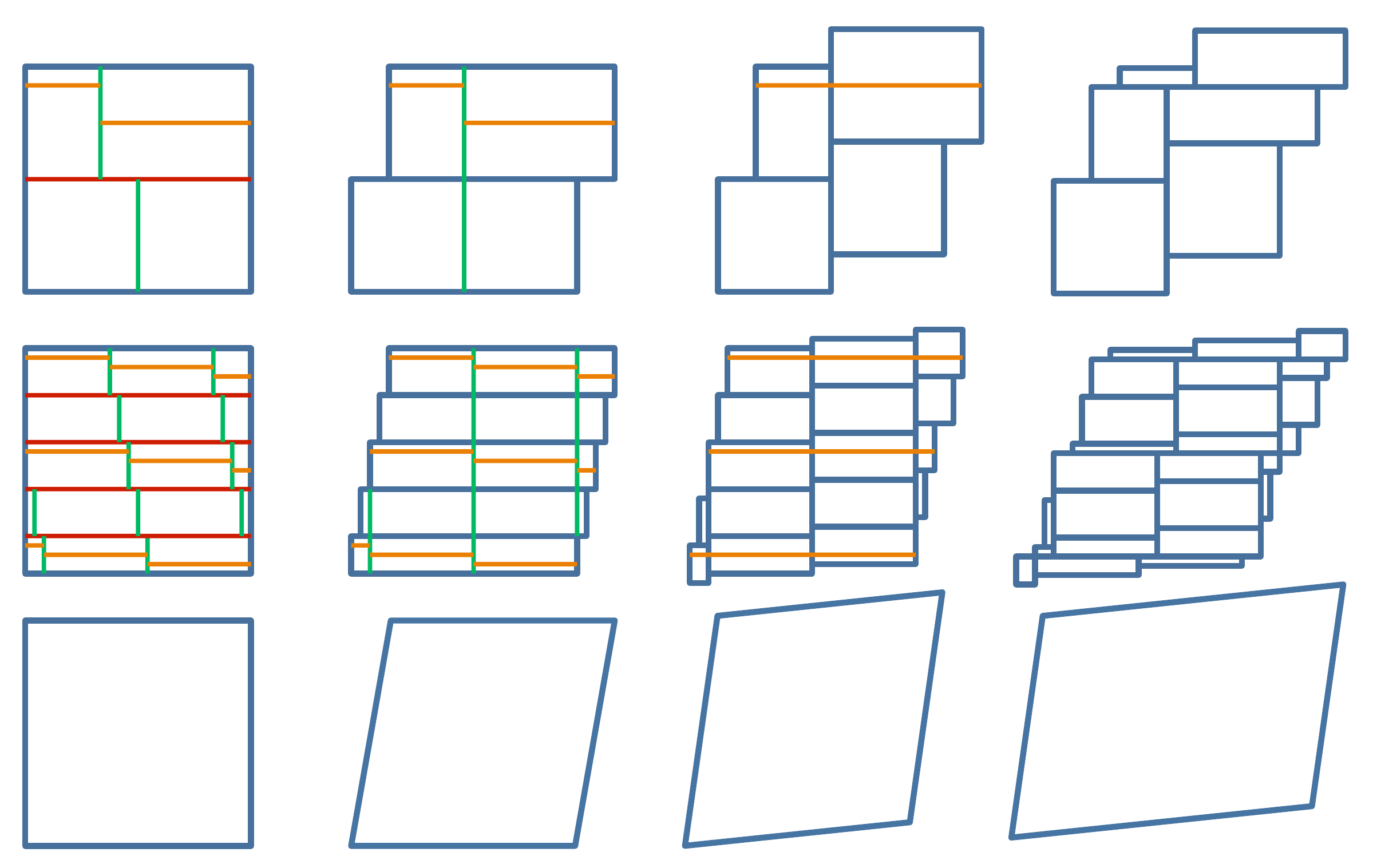}}
    \caption[]{Sequence of deformation induced by three sequential plastic deformations satisfying the scalings $\gamma_1 = \frac{N_{\mathcal{J}_{1\epsilon}} |\mathbf{b}_{1\epsilon}| }{L_2}$, $\gamma_2 = \frac{N_{\mathcal{J}_{2\epsilon}} |\mathbf{b}_{2\epsilon}| }{L_1}$ and $\gamma_3 = \frac{N_{\mathcal{J}_{3\epsilon}} |\mathbf{b}_{3\epsilon}|}{L_2}$. $N_{\mathcal{J}_{1\epsilon}}$, $N_{\mathcal{J}_{2\epsilon}}$ and $N_{\mathcal{J}_{3\epsilon}}$ denote the number of slip lines in each direction, and $\mathbf{b}_{1\epsilon}$, $\mathbf{b}_{2\epsilon}$ and $\mathbf{b}_{3\epsilon}$, the slip on each set of jump sets. The columns left to right show the reference perfect crystal and the deformed configurations after first slip, second and third slip, respectively.}
    \label{Fig:Case5}
\end{center}
\end{figure}

The analysis in Section \ref{Sec:Case3} for two slip systems showed how a precise definition of the plastic deformation tensor required the consideration of the kinks resulting from the pullback of all slip lines. In particular, they were necessary in order to account for the second order terms in the deformation gradient. In this section, we show  how these kinks can also account for higher order terms when multiple plastic deformations are sequentially applied. More specifically, we analyze the composition of three deformation mappings and the corresponding sequence as shown in Fig.~\ref{Fig:Case5}. 

\begin{figure}
\begin{center}
    {\includegraphics[width=0.7\textwidth]{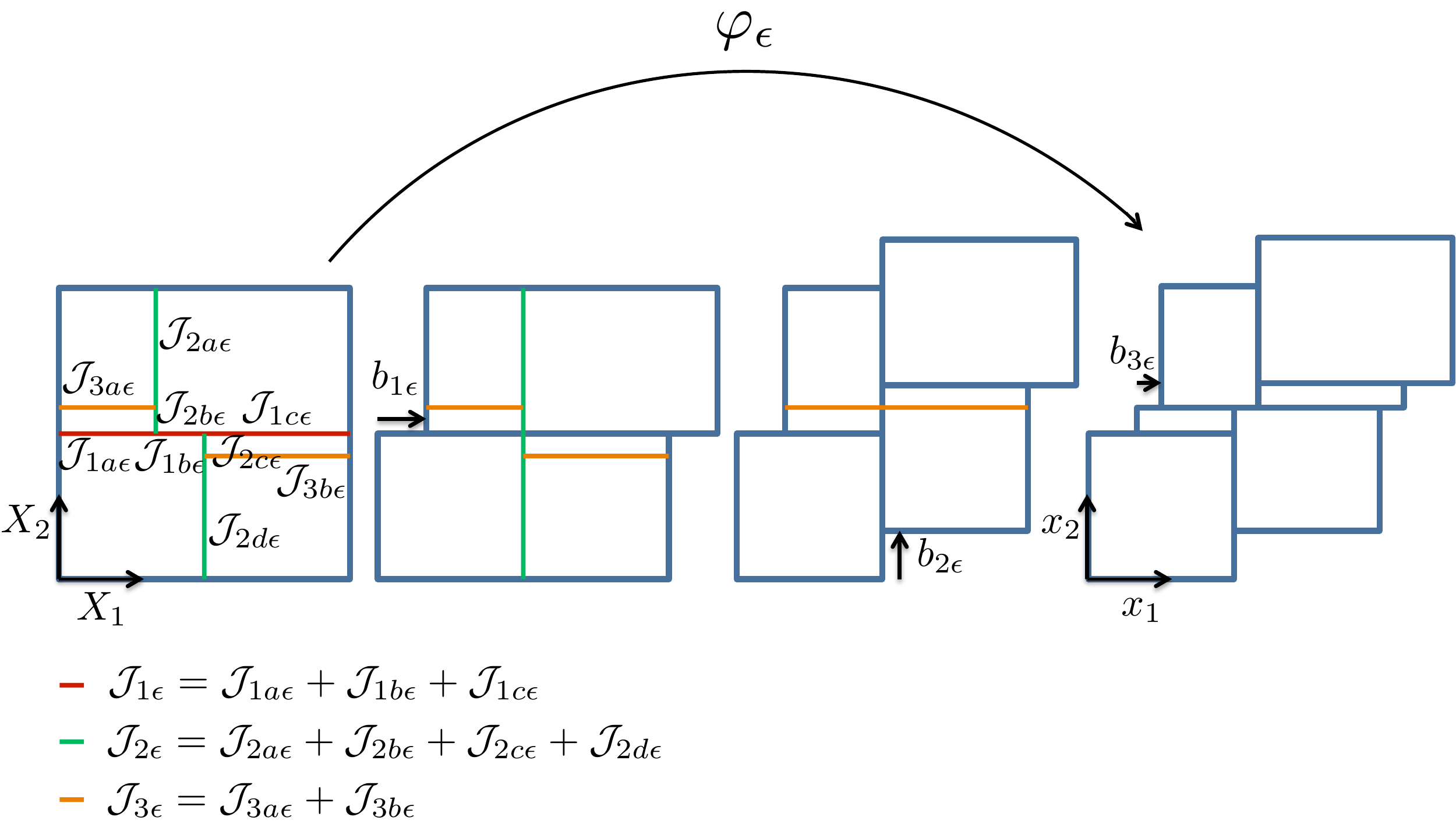}}
    \caption[]{Deformation of a perfect crystal (left image) induced by three sequential plastic deformations (left to right). The third slip line intersects one of the previously formed kinks.}
    \label{Fig:Case5b}
\end{center}
\end{figure}

For each element of the sequence, the displacement jump over the different segments in the reference configuration is, c.f.~Fig.~\ref{Fig:Case5b},
\[ \llbracket \boldsymbol \varphi_{\epsilon} \rrbracket = \llbracket \boldsymbol \varphi_{3\epsilon} \circ \boldsymbol \varphi_{2\epsilon} \circ \boldsymbol \varphi_{1\epsilon} \rrbracket=\left \{ \begin{array} {l l }
\mathbf{b}_{1\epsilon} & \text{on } J_{1a\epsilon} \cup J_{1c\epsilon} \\
\mathbf{b}_{2\epsilon} & \text{on } J_{2a} \cup J_{2d\epsilon} \\
\mathbf{b}_{3\epsilon} & \text{on } J_{3a} \cup J_{3b\epsilon} \\
\mathbf{b}_{1\epsilon} + \mathbf{b}_{2\epsilon} + \mathbf{b}_{3\epsilon} & \text{on } J_{1b\epsilon} \\
\mathbf{b}_{2\epsilon} + \mathbf{b}_{3\epsilon} & \text{on } J_{2b\epsilon} \cup J_{2c\epsilon},
\end{array} \right.
\]
and the deformation gradient (equal to the plastic deformation tensor) can be written as
\begin{equation} \label{Eq:Case5_Fp}
\begin{split}
\mathbf{F}_{\epsilon} = \mathbf{F}^p_{\epsilon} =&\ \mathbf{I}\ \mathcal{L}^2 \\
&+ |\mathbf{b}_{1\epsilon} |\ \mathbf{e}_1\otimes \mathbf{e}_2\ \mathcal{H}^1\lfloor_{\mathcal{J}_{1\epsilon}} + |\mathbf{b}_{2\epsilon} |\ \mathbf{e}_2\otimes \mathbf{e}_1\ \mathcal{H}^1\lfloor_{\mathcal{J}_{2\epsilon}} +  |\mathbf{b}_{3\epsilon} |\ \mathbf{e}_1\otimes \mathbf{e}_2\ \mathcal{H}^1\lfloor_{\mathcal{J}_{3\epsilon}}\\
&+ |\mathbf{b}_{2\epsilon} | \mathbf{e}_2\otimes \mathbf{e}_2\ \mathcal{H}^1\lfloor_{\mathcal{J}_{1b\epsilon}} + |\mathbf{b}_{3\epsilon} |\ \mathbf{e}_1\otimes \mathbf{e}_1\ \mathcal{H}^1\lfloor_{\mathcal{J}_{2b\epsilon}\cup \mathcal{J}_{2c \epsilon}} \\
&+ |\mathbf{b}_{3\epsilon} |\ \mathbf{e}_1\otimes \mathbf{e}_2\ \mathcal{H}^1\lfloor_{\mathcal{J}_{1b\epsilon}},\\
\end{split}
\end{equation}
where $\mathcal{J}_{1\epsilon} = \mathcal{J}_{1a\epsilon} \cup \mathcal{J}_{1b\epsilon} \cup \mathcal{J}_{1c\epsilon}$, $\mathcal{J}_{2\epsilon} = \mathcal{J}_{2a\epsilon} \cup \mathcal{J}_{2b\epsilon}\cup\mathcal{J}_{2c\epsilon} \cup \mathcal{J}_{2d\epsilon}$ and $\mathcal{J}_{3\epsilon} = \mathcal{J}_{3a\epsilon} \cup \mathcal{J}_{3b\epsilon}$. The last term in Eq.~(\ref{Eq:Case5_Fp}) appears when a slip line of $\boldsymbol \varphi_{3\epsilon}$ intersects a previously formed kink, as Fig.~\ref{Fig:Case5b} shows. The sequence satisfies the scaling

\begin{equation}
\begin{split}
&\gamma_1 = \frac{N_{\mathcal{J}_{1\epsilon}} |\mathbf{b}_{1\epsilon}|}{L_2} \\
&\gamma_2 = \frac{N_{\mathcal{J}_{2\epsilon}} |\mathbf{b}_{2\epsilon}|}{L_1} \\
&\gamma_3 = \frac{N_{\mathcal{J}_{3\epsilon}} |\mathbf{b}_{3\epsilon}|}{L_2}, \\
\end{split}
\end{equation}
and converges to a uniform continuous distortion of the form 
\begin{equation} \label{Eq:Case5_limit}
\begin{split}
\mathbf{F}^p L_1 L_2 =&\ \mathbf{I}\ L_1 L_2 \\
&+ N_{\mathcal{J}_{1\epsilon}} L_1 |\mathbf{b}_{1\epsilon}| \ \mathbf{e}_1 \otimes \mathbf{e}_2 + N_{\mathcal{J}_{2\epsilon}} L_2 |\mathbf{b}_{2\epsilon}| \ \mathbf{e}_2 \otimes \mathbf{e}_1+ N_{\mathcal{J}_{3\epsilon}} L_1 |\mathbf{b}_{3\epsilon}| \ \mathbf{e}_1 \otimes \mathbf{e}_2 \\
&+ N_{\mathcal{J}_{1\epsilon}} N_{\mathcal{J}_{2\epsilon}} |\mathbf{b}_{1\epsilon}| |\mathbf{b}_{2\epsilon}| \ \mathbf{e}_2 \otimes \mathbf{e}_2 + N_{\mathcal{J}_{2\epsilon}} N_{\mathcal{J}_{3\epsilon}} |\mathbf{b}_{2\epsilon}| |\mathbf{b}_{3\epsilon}|\ \mathbf{e}_1 \otimes \mathbf{e}_1 \\
&+N_{\mathcal{J}_{1\epsilon}} N_{\mathcal{J}_{2\epsilon}} N_{\mathcal{J}_{3\epsilon}} |\mathbf{b}_{1\epsilon}|\frac{\mathbf{|b}_{2\epsilon}|}{L_2} |\mathbf{b}_{3\epsilon}| \ \mathbf{e}_1 \otimes \mathbf{e}_2 \\
\mathbf{F}^p =& \ \mathbf{I} + \gamma_1 \mathbf{e}_1 \otimes \mathbf{e}_2+ \gamma_2 \mathbf{e}_2 \otimes \mathbf{e}_2+ \gamma_3 \mathbf{e}_1 \otimes \mathbf{e}_2 + \gamma_1 \gamma_2 \mathbf{e}_2 \otimes \mathbf{e}_2 + \gamma_2 \gamma_3 \mathbf{e}_1 \otimes \mathbf{e}_1 \\
&+\gamma_1 \gamma_2 \gamma_3 \mathbf{e}_1 \otimes \mathbf{e}_2,
\end{split}
\end{equation}
where it was taken into account that the probability of forming kinks of the type shown in Fig.  \ref{Fig:Case5b}, is $N_{1\epsilon} N_{2\epsilon}\frac{|\mathbf{b}_{2\epsilon}|}{L_2}$. These kinks in the reference configuration have a length of $|\mathbf{b}_{1\epsilon}|$ and an associated Burgers vector $|\mathbf{b}_{3\epsilon}|$. Equation (\ref{Eq:Case5_limit}) corresponds to the expression that one would expect at the macroscopic scale for the composition of three uniform simple shears
\begin{equation}
\begin{split}
\mathbf{F}^p &= \mathbf{F}^p_3  \mathbf{F}^p_2  \mathbf{F}^p_1 \\
&= \left( \mathbf{I} + \gamma_3 \mathbf{e}_1 \otimes \mathbf{e}_2\right) \left(\mathbf{I} + \gamma_2 \mathbf{e}_2 \otimes \mathbf{e}_2\right) \left( \mathbf{I} + \gamma_1 \mathbf{e}_1 \otimes \mathbf{e}_2\right)  \\
& =  \mathbf{I} + \gamma_1 \mathbf{e}_1 \otimes \mathbf{e}_2+ \gamma_2 \mathbf{e}_2 \otimes \mathbf{e}_2+ \gamma_3 \mathbf{e}_1 \otimes \mathbf{e}_2 \\
&+ \gamma_1 \gamma_2 \mathbf{e}_2 \otimes \mathbf{e}_2 + \gamma_2 \gamma_3 \mathbf{e}_1 \otimes \mathbf{e}_1 \\
&+\gamma_1 \gamma_2 \gamma_3 \mathbf{e}_1 \otimes \mathbf{e}_2.
\end{split}
\end{equation}

Again, for this case, we see that the plastic deformation tensor is uniquely defined as 
\begin{equation} \label{Eq:GeneralFp}
\mathbf{F}^p_{\epsilon} = \mathbf{I}\ \mathcal{L}^2 + \sum_i^{N_{\mathcal{J}_{\epsilon}}} \mathbf{b}_{i\epsilon} \otimes \mathbf{N}_i\ \mathcal{H}^1 \lfloor_{\mathcal{J}_{i\epsilon}},
\end{equation}
where the sum occurs over the pullback of all slip lines to the reference configuration and therefore it includes kinks induced by the pullback. The Burgers vector associated to the kinks is in general not orthogonal to the normal of the slip segment ($\mathbf{b}_{i\epsilon}\cdot \mathbf{N}_i\neq 0$).

\subsection{General elastoplastic deformations} \label{Sec:GeneralEP}
The previous examples were exclusively concerned with compatible elastic and plastic deformations. For these cases, the kinematic analyses were sufficient to uniquely characterize the plastic deformation tensor as a function of the slip on the jump set, c.f. Eq. (\ref{Eq:GeneralFp}). The range of applicability of this definition can be extended to the incompatible setting where dislocations are present in the system. This extension can be physically justified in two steps. On the one hand, in any subdomain free of dislocations, the above definition of the plastic deformation is exact as argued for domains with compatible deformations. On the other hand, the volume associated to the neighborhood of the dislocations vanishes in the continuum limit and their contribution to the plastic deformation tensor can also be proven to vanish. More specifically, a scaling for the number of dislocations ($N_{\epsilon}$) of the form
\begin{equation}
N_{\epsilon} \epsilon < C
\end{equation}
allows physically observed distribution of dislocations, for instance, dislocation walls and pileups or homogenous distributions of dislocations. The contribution of the dislocation cores to the plastic deformation tensor according to this scaling is of the order $N_{\epsilon}\epsilon^2 < C \epsilon$, which vanishes in the limit as $\epsilon$ tends to zero. 

The discrete definitions of the deformation tensors $\mathbf{F}_{\epsilon}, \mathbf{F}^e_{\epsilon}$ and $\mathbf{F}^p_{\epsilon}$ given by Eqs. (\ref{Eq:Definition_F}), (\ref{Eq:Definition_Fe}) and (\ref{Eq:GeneralFp}), can therefore be considered valid for arbitrary elastoplastic deformations in the presence of dislocations. However, the proof for the multiplicative decomposition of the total deformation gradient in this general setting is far for being trivial. Indeed, the limit of a product of two weakly converging tensors ($\mathbf{F}^e_{\epsilon}$ and $\mathbf{F}^p_{\epsilon}$ in this case) do not correspond in general to the product of the limit of the individual tensors. In the compatible case,  the elastic deformation tensor could be taken as constant along the sequence highly simplifying the proof. In the presence of dislocations, the coupling between the elastic and the plastic deformation prevents that assumption. A general proof for $\mathbf{F}=\mathbf{F}^e\mathbf{F}^p$ in the continuum limit with the definitions provided here requires energetic considerations and involved mathematical details. This proof is currently in progress by the authors and is out of the scope of the present paper, which is purely based on kinematic considerations.

\section{Discrete dislocation density tensor} \label{DDTensor}
Dislocations can be regarded as the boundary of the active slip surfaces. They are characterized by their orientation (in three dimensions) and their Burgers vector, which can be computed as the sum of the Burgers vector of the slip lines that terminate at each dislocation. Both of these quantities, when referred to the reference configuration, are contained in the definition of the discrete plastic deformation tensor

\begin{equation} \label{Eq:Fp_definition}
\mathbf{F}^p_{\epsilon} = \mathbf{I} \mathcal{L}^2 + \sum_i^{N_{\mathcal{J}_{\epsilon}}} \mathbf{b}_{i\epsilon} \otimes \mathbf{N}_i \mathcal{H}^1 \lfloor_{\mathcal{J}_{i\epsilon}}.
\end{equation}
In particular, Curl $\mathbf{F}^p_{\epsilon}$, computed in a distributional sense, exactly delivers the discrete version of the dislocation density tensor $\mathbf{G}_{\epsilon}$
\begin{equation} \label{Eq:Curl_Fp}
 \text{Curl}\ \mathbf{F}^p_{\epsilon} = \sum_i^{N_{\mathcal{J}_i}} \mathbf{b}_{i\epsilon} \left(\delta_{\mathbf{X}_{i1}}-\delta_{\mathbf{X}_{i2}} \right) = \sum_j^{N_{\epsilon}} \mathbf{b}_{j\epsilon} \delta_{\mathbf{X}_j} = \mathbf{G}_{\epsilon},
\end{equation}
where $\mathbf{X}_{i1}$ and $\mathbf{X}_{i2}$ are the two endpoints of the slip segment $\mathcal{J}_i$, defined such that $\left(\mathbf{X}_{i1}-\mathbf{X}_{i2}\right)\cdot \mathbf{N}_i > 0$; and $\mathbf{X}_j$ is the position of dislocation $j$ with associated Burgers vector $\mathbf{b}_{j\epsilon}$. The proof for Eq. (\ref{Eq:Curl_Fp}) is given in \ref{Sec:AppendixB} and takes into account the fact that $\mathbf{b}_{j\epsilon}$ is constant in magnitude and direction for each slip segment $\mathcal{J}_i$. 

The direct connection between Curl $\mathbf{F}^p_{\epsilon}$ and the dislocation density tensor $\mathbf{G}_{\epsilon}$ was possible thanks to two considerations: (i) a reference configuration that consisted on a perfect (defect free and not elastically deformed) crystalline structure; and (ii) the translational periodicity of the lattice, which restricted the slip lines to be straight. In the absence of either of the two conditions, the two measures would not coincide. Indeed, an elastically deformed reference configuration would induce non-constant Burgers vectors along each slip line, and Curl $\mathbf{F}^p_{\epsilon}$ would be a measure distributed along the jump set rather than concentrated at its endpoints. Equivalently, if slip lines were permitted to be curved in the reference configuration, one could construct a circular slip line which results from rotating the inner circle with respect to the outer domain. This specific example would have a non-vanishing associated \textit{Curl} (maintaining the definition for the plastic deformation tensor), whereas the dislocation content and the elastic energy would in principle be null.

The continuum limit of Eq. (\ref{Eq:Curl_Fp}) has an easy mathematical justification. In particular, convergence of the plastic deformation tensor and the dislocation density tensor\footnote{convergence of these two quantities is ensured by the scaling $(|\mathcal{J_{\epsilon}}|+N_{\epsilon})\epsilon < C$.}
\begin{equation}
\begin{split}
&\mathbf{F}^p_{\epsilon} \overset{*}{\rightharpoonup} \mathbf{F}^p \\
&\mathbf{G}_{\epsilon}\overset{*}{\rightharpoonup} \mathbf{G} \\
\end{split}
\end{equation}
guarantees the following equivalence
\begin{equation}
\mathbf{G} = \text{Curl }\mathbf{F}^p.
\end{equation}

That is, $\text{Curl }\mathbf{F}^p$ exactly provides a measure for the net Burgers vector (measured in the reference configuration) per unit area of the reference configuration, regardless of the number of activated slip systems. This result was obtained with the standard notions of weak derivatives and does not require the use of Burgers circuit in any artificial intermediate configuration.

\subsubsection{Physical interpretation of the \textit{Curl} operator} \label{Sec:CurlUnderstanding}

The micromechanical understanding of $\mathbf{F}^p_{\epsilon}$ and its \textit{Curl} provides the necessary elements to physically interpret the \textit{Curl} operator over a composition of plastic deformations. To that end, consider the example of Sec. \ref{Sec:Case3} and define the following tensors
\begin{equation}
\begin{split}
&\mathbf{F}^p_{\epsilon} = \mathbf{I}\ \mathcal{L}^2 + |\mathbf{b}_{1\epsilon} |\ \mathbf{e}_1\otimes \mathbf{e}_2\ \mathcal{H}^1\lfloor_{\mathcal{J}_{1\epsilon}} + |\mathbf{b}_{2\epsilon} |\ \mathbf{e}_2\otimes \mathbf{e}_1\ \mathcal{H}^1\lfloor_{\mathcal{J}_{2\epsilon}} +  |\mathbf{b}_{2\epsilon} | \mathbf{e}_2\otimes \mathbf{e}_2\ \mathcal{H}^1\lfloor_{\mathcal{J}_{1b\epsilon}}  \\
&\mathbf{F}^p_{1\epsilon} =  \mathbf{I}\ \mathcal{L}^2 + |\mathbf{b}_{1\epsilon} |\ \mathbf{e}_1\otimes \mathbf{e}_2\ \mathcal{H}^1\lfloor_{\mathcal{J}_{1\epsilon}} \\
&\mathbf{F}^p_{2\epsilon} = \mathbf{I}\ \mathcal{L}^2 + |\mathbf{b}_{2\epsilon} |\ \mathbf{e}_2\otimes \mathbf{e}_1\ \mathcal{H}^1\lfloor_{\mathcal{J}_{2\epsilon}}.  \\
\end{split}
\end{equation}

\begin{figure}
\begin{center}
    {\includegraphics[width=0.4\textwidth]{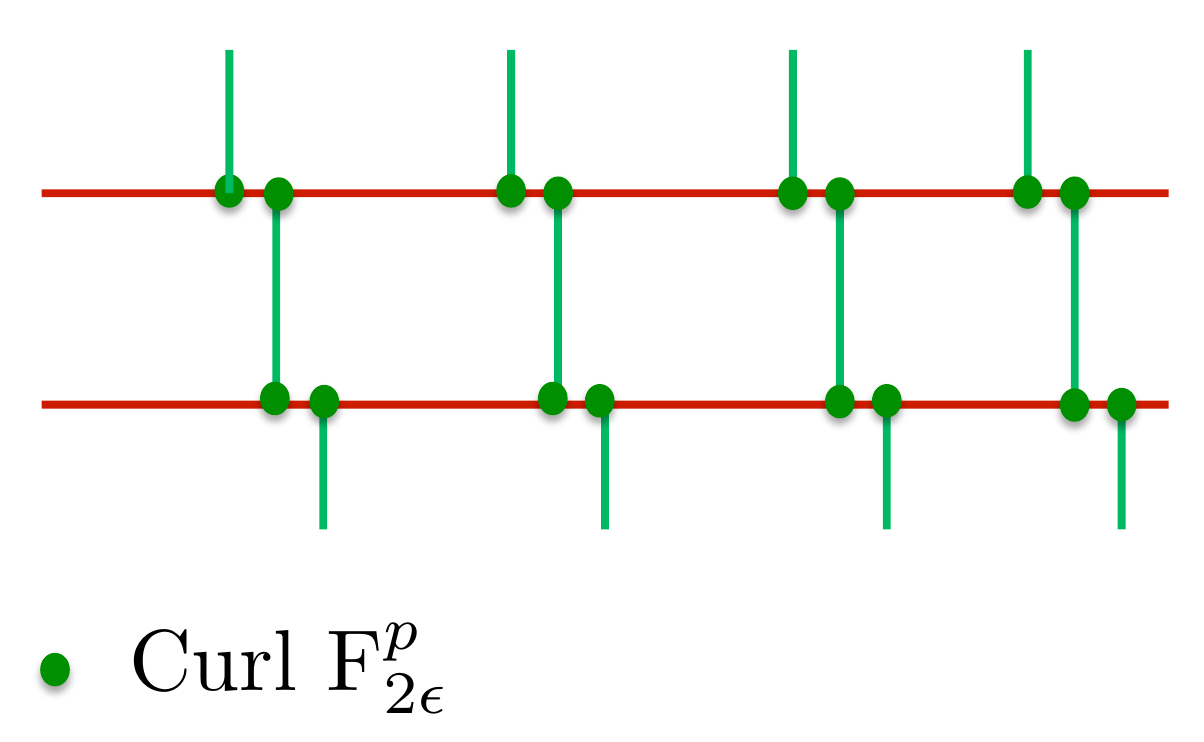}}
    \caption[]{Distributional \textit{Curl} associated the vertical slip lines.}
    \label{Fig:Case3c}
\end{center}
\end{figure}

In general, the points of intersection between the different segments of the jump set correspond to dislocations if the net Burgers vector (sum of the Burgers vector of each segment terminating at the node) does not vanish. For the given example, it is then immediate that  Curl $\mathbf{F}^p_{\epsilon}$ and Curl  $\mathbf{F}^p_{1\epsilon} $ vanish, as physically expected for the compatible deformations $\boldsymbol \varphi_{\epsilon}$ and $\boldsymbol \varphi_{1\epsilon}$. However, their difference does not correspond to Curl $\mathbf{F}^p_{2\epsilon}$, as one might initially suspect, c.f. Fig.\ref {Fig:Case3c}. Rather, 
\begin{equation}\label{Eq:Curl}
 \text{Curl } \mathbf{F}^p_{\epsilon} = 0=  \text{Curl } \mathbf{F}^p_{2\epsilon} + \text{Curl } \left(  |\mathbf{b}_{2\epsilon} | \mathbf{e}_2\otimes \mathbf{e}_2\ \mathcal{H}^1\lfloor_{\mathcal{J}_{1b\epsilon}} \right),
\end{equation}
where the last term corresponds to the kinks and is induced by the pullback operation. Physically, the sum of the last two terms measures the dislocations introduced by the second deformation mapping (in this case, none), or equivalently, $\tilde{\text{C}}\text{url}\  \tilde{\mathbf{F}}^{p}_2 =0$ with $\tilde{\mathbf{F}}^p_{2\epsilon}=|\mathbf{b}_{2\epsilon}|\ \tilde{\mathbf{e}}_2\otimes \tilde{\mathbf{e}}_1 \mathcal{H}^1 \lfloor_{\tilde{\mathcal{J}_{2\epsilon}}}$. 

The macroscopic analog of Eq. (\ref{Eq:Curl}) is given by
\begin{equation}
\begin{split}
[ \text{Curl } \left( \tilde{\mathbf{F}}^{p}_2 \mathbf{F}^{p}_1 \right) ]_{\tilde{I}}  &=  \tilde{\mathbf{F}}^{p}_{2 \tilde{I} \tilde{N}} \mathbf{F}^p_{1\tilde{N}M,K} e_{3KM} + \tilde{\mathbf{F}}^{p}_{2 \tilde{I} \tilde{N},K} \mathbf{F}^p_{1\tilde{N}M} e_{3KM} \\
 0_{\tilde{I}} &= \tilde{\mathbf{F}}^{p}_{2 \tilde{I} \tilde{N},K} \mathbf{F}^p_{1\tilde{N}M} e_{3KM}, \\
\end{split}
\end{equation}
where the last equation might not seem intuitive at first. However, by applying the chain rule (recall that $\boldsymbol \varphi_1$ is compatible in this specific example)
\begin{equation} \label{Eq:Curl_Fp2Fp1}
\begin{split}
\tilde{\mathbf{F}}^{p}_{2 \tilde{I} \tilde{N},K} \mathbf{F}^p_{1\tilde{N}M} e_{3KM} &= \tilde{\mathbf{F}}^{p}_{2 \tilde{I} \tilde{N},\tilde{P}} \mathbf{F}^p_{1\tilde{P}K} \mathbf{F}^p_{1\tilde{N}M} e_{3KM} \\
& = \tilde{\mathbf{F}}^{p}_{2 \tilde{I} \tilde{N},\tilde{P}} \det \mathbf{F}^p_1 e_{3\tilde{P}\tilde{N}} \\
& = [\tilde{\text{C}}\text{url } \tilde{\mathbf{F}}^p_{2} ]_{\tilde{I}}.
\end{split}
\end{equation}

The same expression is then recovered at the discrete and continuum scale. This analysis, can be easily extended to the case where the second deformation mapping introduces some dislocations. However, the pullback operation done in Eq. (\ref{Eq:Curl_Fp2Fp1}) requires that the first plastic deformation is dislocation free.

\section{Discussion} \label{Sec::Additional}
\subsection{Existence of $\mathbf{F}=\mathbf{F}^e\mathbf{F}^p$} \label{Sec:Existence}

The two-dimensional kinematic analyses performed at the microscale suggested
physically-sound definitions for the different deformation tensors
$\mathbf{F}_{\epsilon}$, $\mathbf{F}^e_{\epsilon}$ and
$\mathbf{F}^p_{\epsilon}$. It was then shown for several dislocation-free
examples, that these definitions lead in the continuum limit to the well-known
multiplicative decomposition of the deformation gradient,
i.e. $\mathbf{F}=\mathbf{F}^e\mathbf{F}^p$. The main elements of the
treatment of  general elastoplastic deformations with dislocations were sketched out in Section \ref{Sec:GeneralEP} and its complete version is currently in progress. Under the hypothesis of the semi-continuous model and the scaling considered, it is therefore expected that the multiplicative decomposition of the deformation gradient exists, even in situations where, at the macroscopic scale, a stress-free configuration is inaccessible. The key point is that the definitions for  $\mathbf{F}^e$ and $\mathbf{F}^p$ come from their discrete counterpart, which do not resort to any artificial intermediate configuration, nor they required the use of a macroscopic unloading path for their definition.

The reference configuration used to characterize the deformation was assumed
to be a perfect unstressed crystal. However, the previous conclusions can be
extended to general unrelaxed reference configurations, as long as a complete
pullback of all the slip lines to the chosen reference is performed in order
to define $\mathbf{F}^p_{\epsilon}$. This would generally lead to curved slip
lines with non-constant Burgers vectors associated to them, increasing the
complexity of the kinematic analyses. For the sake of simplicity a perfect
crystal is here considered for the reference configuration. 

\subsection{Uniqueness of $\mathbf{F}=\mathbf{F}^e\mathbf{F}^p$}

The question of uniqueness of  the multiplicative decomposition and the
associated invariance requirements for the formulation of physical
constitutive laws has been debated in the literature \citep{LubardaLee1981,
  CaseyNaghdi1980, CaseyNaghdi1981, Dashner1986, GreenNaghdi1971,
  WeberAnand1990, Gupta2007}. The micromechanical perspective adopted here
makes the definition of $\mathbf{F}^e$ and $\mathbf{F}^p$ unique
  from the microstructure, and the invariance requirements
unambiguous. Defining by superscript * the fields after a superposed rotation
$\mathbf{Q}$, it is then easy to see that our definitions lead to
\begin{equation} \label{Eq:InvarianceRequirements}
\begin{split}
&\mathbf{F}^* = \mathbf{Q F} \\
&\mathbf{F}^{e*} = \mathbf{Q F}^e \\
&\mathbf{F}^{p*} = \mathbf{F}^p,\\
\end{split}
\end{equation}
as it is typically assumed. More precisely, Eqs.~(\ref{Eq:Fp_definition}) and (\ref{Eq:Definition_Fe}) indicate that the plastic deformation tensor remains unchanged under a superposed rigid body motion and that the elastic deformation includes the rigid rotation of the body, consistently with Eq.~(\ref{Eq:InvarianceRequirements}).

The uniqueness of $\mathbf{F}^p$ is also essential when defining a dislocation density tensor as a function of that measure. If $\mathbf{F}^p$ were to be defined up to a rotation as it is sometimes stated, the dislocation density tensor would be ill-defined. In particular, for $\mathbf{Q} \in SO(2)$ a non-uniform rotation, the dislocation density $\text{Curl} (\mathbf{Q F}^p)$
\begin{equation}
[\text{Curl} (Q F^p)]_I = (Q F^P)_{IJ,M} e_{3MJ} = [Q \text{Curl} F^p ]_{I} +Q_{IN,M} F^p_{NJ} e_{3MJ}
\end{equation}
 would vary according to the arbitrariness of the chosen field $\mathbf{Q}$. The definition of $\mathbf{F}^p_{\epsilon}$ given by Eq. (\ref{Eq:Fp_definition}) prevents this unphysical consequence when considering the associated dislocation density tensor.

\subsection{Micromechanical understanding of the plastic deformation tensor.}

The macroscopic plastic deformation tensor $\mathbf{F}^p$ can be seen as the continuum limit of its discrete counterpart $\mathbf{F}^p_{\epsilon}$
\begin{equation} \label{Eq:GeneralFp2}
\mathbf{F}^p_{\epsilon} = \mathbf{I}\ \mathcal{L}^2 + \sum_i^{N_{\mathcal{J}_{\epsilon}}} \mathbf{b}_{i\epsilon} \otimes \mathbf{N}_i\ \mathcal{H}^1 \lfloor_{J_{i\epsilon}},
\end{equation}
which is defined as a sum over the pullback of the slip segments to the reference configuration, assumed stress-free. This slip structure $\mathcal{J}$ is in general very complex and includes kinks induced by the pullback operation. Despite the small length of these kinks, they become essential for capturing the higher order terms in macroscopic deformation that involve multiple slip systems, and is responsible for the non-commutative character of deformations in the finite kinematic setting. An important point from this perspective is that $\mathbf{F}^p$ is not defined via a macroscopic unloading path, which is in general not possible. Rather, it is defined as the weak limit (or average) of the discrete analog. It is only at this discrete level that physical local relaxations are well defined almost everywhere.

\begin{figure}
\begin{center}
    {\includegraphics[width=0.8\textwidth]{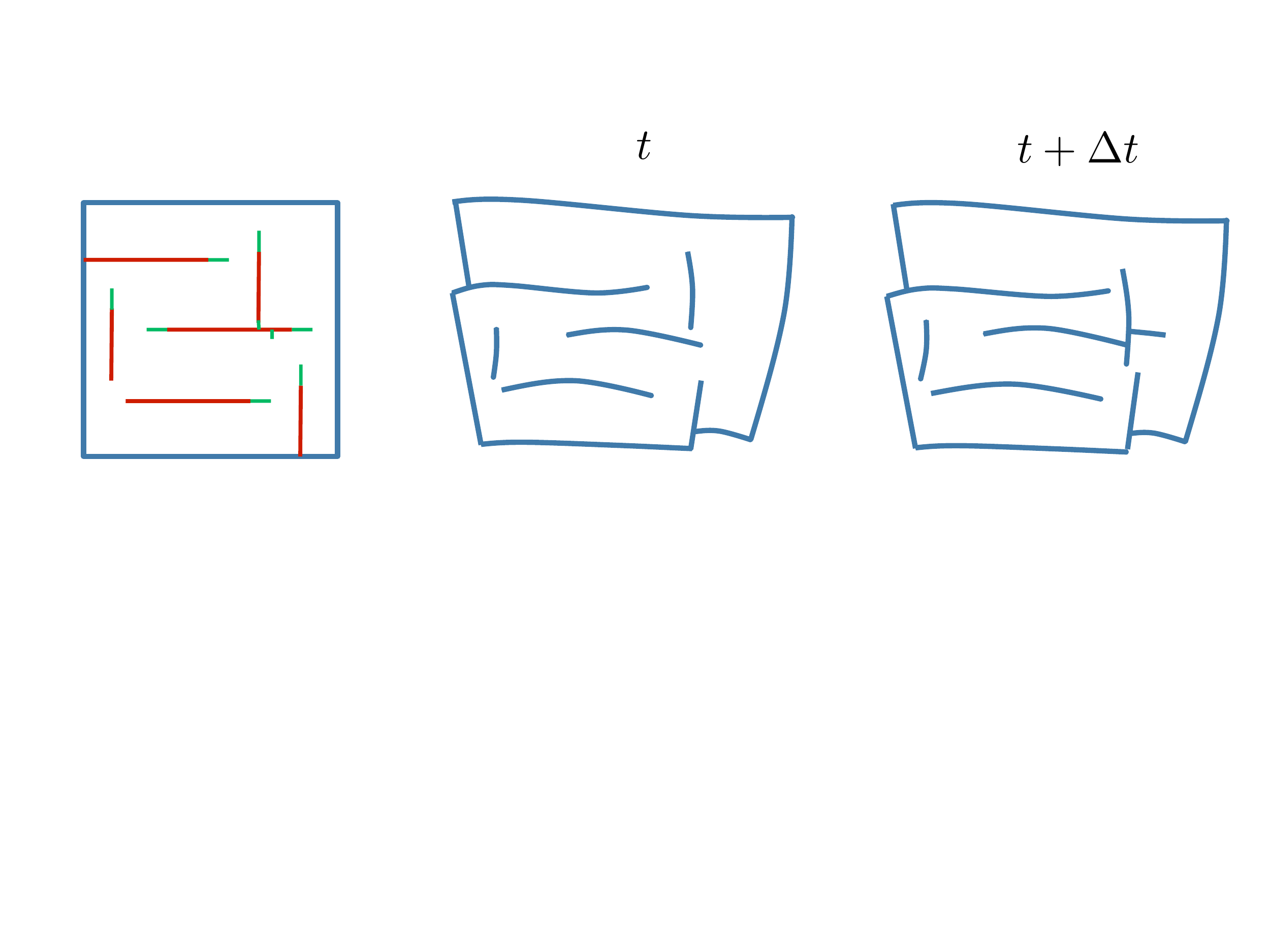}}
    \caption[]{Elastoplastic deformation induced by two slip systems at time $t$ and $t+\Delta t$. The pullback of the jump set at time $t$ is shown in red in the reference configuration (left image); and the additional jump set between $t$ and $t+\Delta t$ is represented in green. }
    \label{Fig:Incremental}
\end{center}
\end{figure}

A notable characteristic of Eq. (\ref{Eq:GeneralFp2}) is that the relation between the total plastic deformation tensor $\mathbf{F}^p_{\epsilon}$ and the underlying dislocations and slip structure is defined via an explicit equation. This is in contrast to the classical relation, c.f. \citep{Rice1971}
\begin{equation} \label{Eq:FpRate2}
\dot{\mathbf{F}}^p \mathbf{F}^{p-1} = \sum_k \dot{\gamma}_k \mathbf{M}_k \otimes \mathbf{N}_k,
\end{equation}
that relates $\mathbf{F}^p$ and the plastic slip. In Eq. (\ref{Eq:FpRate2}) $\dot{\gamma}_k$ is the strain rate on direction $\mathbf{M}_k$ of a slip system with normal $\mathbf{N}_k$. Both equations, (\ref{Eq:GeneralFp2}) and (\ref{Eq:FpRate2}), are actually in agreement under suitable assumptions. Without loss of generality, consider the plastic deformation induced by two slip systems ($N_s = 2$) at times $t$ and $t+ \Delta t$, c.f Fig. \ref{Fig:Incremental}. From the example in Section \ref{Sec:Case2}, it is suggested that the limit and the product commute for the types of deformations here considered\footnote{A rigorous proof of this statement requires energetic considerations, which are out of the scope of this paper.}. As a result
\begin{equation} \label{Eq:Increment_Fp}
\mathbf{F}^p_{t+\Delta t} = \lim_{\epsilon \rightarrow 0} \left(\mathbf{I}\ \mathcal{L}^2 + \sum_i^{N_{\Delta \mathcal{J}}} \mathbf{b}_{i\epsilon} \otimes \mathbf{N}_i \mathcal{H}^1\lfloor_{\Delta \mathcal{J}_i}\right) \mathbf{F}^p_t,
\end{equation}
where $\Delta \mathcal{J}_i$ is the new jump set in the reference configuration created between times $t$ and $t+\Delta t$, c.f. Fig. \ref{Fig:Incremental}. Those could be generated by the glide of existing dislocations, or by the creation of new segments and the corresponding dislocations. Under the scaling 
\begin{equation}
\Delta \gamma_k = \frac{ \sum_{ \{i \in \text{system } k \}} |\Delta \mathcal{J}_i| |\mathbf{b}_{i\epsilon}|}{|\Omega|},
\end{equation}
Eq. (\ref{Eq:Increment_Fp}) can be written as
\begin{equation}
\mathbf{F}^p_{t+\Delta t} = \Big [\mathbf{I} + \sum_k^{N_s = 2} \Delta \gamma_k \mathbf{M}_k \otimes \mathbf{N}_k + \mathcal{O}(\Delta \gamma)^2 \Big] \mathbf{F}^p_t. 
\end{equation}

The resulting rate of plastic deformation then reads
\begin{equation}
\dot{\mathbf{F}}^p = \lim_{\Delta t \rightarrow 0} \frac{\mathbf{F}^p_{t+\Delta t}- \mathbf{F}^p_t}{\Delta t} = \left( \sum_k \lim_{\Delta t \rightarrow 0} \frac{\Delta \gamma_k}{\Delta t}  \mathbf{M}_k \otimes \mathbf{N}_k \right) \mathbf{F}^p,
\end{equation}
where $\mathbf{M}_k$ is the direction of the Burgers vector of slip system $k$, with $k=1, ... N_s$ ($N_s = 2$ in this case). Equivalently,
\begin{equation}\label{eqdotfpfpmn}
\dot{\mathbf{F}}^p \mathbf{F}^{p-1} = \sum_k \dot{\gamma}_k \mathbf{M}_k \otimes \mathbf{N}_k,
\end{equation}
thus recovering the well-known relation.

In summary, we have shown that the simultaneous and gradual activity of several
slip systems implied by Eq.~(\ref{Eq:FpRate2}) can be recovered in the current
framework based on Eq.~(\ref{Eq:GeneralFp2}) if the total deformation is discretized in time 
into a large number of small deformations. Activity along each of the slip systems is thus decomposed
into many small steps along individual segments, which result from the dislocations motion 
in the corresponding time interval. These small segments arising from different systems
are then composed with each other. Since small slips commute, the (arbitrary) order of
the small steps is irrelevant and Eq.~(\ref{Eq:FpRate2}) is recovered in the limit.

As an additional remark, Eq. (\ref{Eq:GeneralFp2}) provides a physical meaning for the use of indices related to the same reference frame for the components of $\mathbf{F}^p$ when expressed with indicial notation ($F^p_{IJ}$). Equivalently, the hybrid character of the macroscopic elastic deformation tensor remains meaningful ($F^e_{iJ}$).

\subsection{Dislocation density tensor} \label{Sec:DisDenTens}

The dislocation density measure $\mathbf{G}$ in two dimensions delivers the net Burgers vector per unit area. Different reference frames can be chosen to express those quantities.  In this work, we used the reference configuration as a natural space for both the Burgers vector and the surface area. It was then obtained at the discrete and continuum level, that the dislocation density is directly related to the \text{Curl} of the plastic deformation tensor
\begin{equation} \label{Eq:DDTensor_discussions}
\begin{split}
&\mathbf{G}_{\epsilon} =\text{Curl } \mathbf{F}^p_{\epsilon} \\
&\mathbf{G} =\text{Curl } \mathbf{F}^p. \\
\end{split}
\end{equation}

The additive structure of the obtained expression for $\mathbf{F}^p$, c.f.  Eq. (\ref{Eq:GeneralFp2}), and its direct relation (not in rate form) to the individual slips, allows a direct computation of the contribution of each slip system to the total incompatibility Curl $\mathbf{F}^p$.

The equivalence between the dislocation density tensor $\mathbf{G}$ and Curl
$\mathbf{F}^p$  is contingent to having a stress-free reference frame. For a
deformed reference configuration, the elastoplastic decomposition of the
deformation gradient is expected to remain valid as discussed in Section
\ref{Sec:Existence}. However, Eq. (\ref{Eq:DDTensor_discussions}) would not
hold in that case. 

\subsection{Three dimensional setting} \label{Sec:3D}

The two-dimensional analyses previously performed are suggestive of the following expression for the discrete plastic deformation tensor in three dimensions
\begin{equation}
\mathbf{F}^p_{\epsilon} = \mathbf{I}\ \mathcal{L}^3 + \sum_i^{N_{\mathcal{J\epsilon}}} \mathbf{b}_{i\epsilon} \otimes \mathbf{N}_i\ \mathcal{H}^2 \lfloor_{J_{i\epsilon}}, \\
\end{equation}
where the jump set $\mathcal{J}$ is now composed of the union of planar surfaces. The dislocations in a three dimensional space can also be regarded as the lines bounding the slip surfaces, and their location and intensity can be obtained via the distributional Curl of $\mathbf{F}^p_{\epsilon} $, leading to
\begin{equation}
\mathbf{G}_{\epsilon} = \text{Curl } \mathbf{F}^p_{\epsilon} = \sum_i^{N_{\mathcal{J}_{\epsilon}}} \mathbf{b}_{i\epsilon} \otimes \mathbf{\xi}_i\ \mathcal{H}^1 \lfloor_{\partial \mathcal{J}_{i\epsilon} },\\
\end{equation}
where $\boldsymbol \xi$ is a vector tangent to the dislocation and obeying the right-hand rule with the normal $\mathbf{N}_i$ of each surface element $\mathcal{J}_i$.  Similarly to the two-dimensional case, this direct relation between the plastic deformation tensor and the dislocation content is contingent to the use of a stress-free reference configuration.

The continuum counterpart of Eq. (\ref{Eq:3D_G}) is then immediately given by
\begin{equation} \label{Eq:3D_G}
\mathbf{G} = \text{Curl } \mathbf{F}^p,
\end{equation}
which delivers, as expected, the well-known conservation rule for the Burgers vector
\begin{equation}
\int_{\omega} \nabla \cdot \mathbf{G} = \int_{\partial \omega} \mathbf{G} \cdot \mathbf{N} =0
\end{equation}
This equation implies that all dislocations that enter a subdomain $\omega \subset \Omega$ necessarily exit it. Equivalently, dislocations are either closed lines in the interior of  $\Omega$, or exit its boundary.

\subsection{Relation to previous studies}
\label{secpreviousstudies}

The additive decomposition of the (weak) deformation gradient into an
absolutely continuous part and a singular part similar to Eq.~(\ref{Eq:Fp_Case1})
has long been used in the
kinematic characterization of elasto-plasticity in the linearized kinematic
setting, 
leading to the definition and study of the space of functions of bounded
deformation \citep{TemamStrang1980,AmbrosioCosciaDalmaso1997}. 
Using this decomposition one can easily devise models where the 
elastic mechanism corresponds to the absolutely continuous part of the
deformation gradient, and the plastic mechanism to the singular part which is
concentrated on slip surfaces. After homogenization this 
leads naturally  to an additive decomposition of the deformation
gradient in the linear setting ($ \boldsymbol \varepsilon  =\boldsymbol \varepsilon^e  +\boldsymbol \varepsilon^p $).

Additive decompositions of the same sort are also used in the theory of structured deformations, introduced by
\cite{DelpieroOwen1993}, 
where deformations are considered
piecewise continuous and therefore permit plastic slip and the opening of
cracks. For a review of the developments of this theory 
we refer to the recent book \citep{DelpieroOwenBook2004}.
 In such body of work, the authors adopt a \emph{top-down}
approach, in which they study the existence of disarrangements over a set of
crystallographic slip systems that could approximate a given singular part of
the deformation gradient and  establish invariance requirements for
constitutive models. Approximations are found in the limit of small
deformations, which are sufficient to rigorously derive an expression analog
to Eq.~(\ref{Eq:FpRate2}) characterizing the evolution of the plastic
deformation 
tensor in  fcc and bcc crystals \citep{Owen2002}.
Analogies are also drawn to the multiplicative
decomposition of the 
deformation gradient via suitable rearrangements of the additive
decomposition, which are argued in terms of pull back and push forward of the
different tensors involved.  
In particular, the basic additive decomposition
$\mathbf{F}=\mathbf{G}+\mathbf{M}$  proposed in 
\cite{DelpieroOwen1993}, where $\mathbf{G}$ is invertible, can lead to a
multiplicative decompositions both of the type 
$\mathbf{F}^p\mathbf{F}^e$ and $\mathbf{F}^e\mathbf{F}^p$ by factorizing
$\mathbf{G}$ either to the right or to the left in the additive decomposition
respectively.

In contrast, the approach taken in this work could be characterized as
\emph{bottom-up}.  We study the physics of slip and the associated geometric
constraints on the kinematics, to physically define unique
deformation tensors $\mathbf{F}$, $\mathbf{F}^e$ and $\mathbf{F}^p$ from the microstructure. 
Our  definitions encompass
careful treatment of intersecting slip systems and are valid for small and
large deformations. Based on these microstructural definitions\footnote{The
  constraints that slip imposes on the jump of the deformation mapping results
  essential in this derivation, and appears to be novel in the mathematics
  literature}, the multiplicative decomposition of the deformation gradient is
recovered in the continuum limit without its a prori assumption.  Furthermore, dislocations are
explicitly treated in this bottom-up approach, delivering a physically-based
and mathematically-consistent definition for the dislocation density tensor,
and providing with an identical formula to that of Rice, c.f.~Eq.~(\ref{Eq:FpRate2}), for the evolution of
the plastic deformation tensor. 

The kinematic discussion in the present paper forms the basis for a future
energetic modeling. 
The use of relaxation theory to study crystal plasticity 
using functionals with linear growth defined on Sobolev
functions has been very successfull  for geometrically linear models, see for example 
\citep{BraidesDefranceschiVitali1997,ContiOrtiz05}, and was recently extended
to geometrically nonlinear models
 in \citep{OrtizRepetto1999,ContiDolzmannKreisbeck2013b}. 
In the context of structured deformations, energetic 
models have already been proposed and studied for example in
\cite{ChoksiDelpieroFonsecaOwen1999,DeseriOwen2000,DeseriOwen2002}, see 
also \citep{DelpieroOwenBook2004}. The theory of
relaxation in the context of structured deformations was discussed starting in
\cite{ChoksiFonseca1997}; see \citep{BaiaSantosMiguel2012} for recent
developments. 

\section{Summary and concluding remarks} \label{Sec::Conclusion}
Due to the large deformations incurred by ductile materials prior to failure, the finite kinematic framework is of necessity for macroscopic modeling of plasticity. However, many issues related to its formulation in the continuum setting remain topics of ongoing discussion. Some of the examples treated in this paper include the decomposition of the macroscopic deformation gradient $\mathbf{F} =\mathbf{F}^e \mathbf{F}^p$, the definition of the elastic and plastic deformation tensor or the measure of the dislocation density content in the body.

Some of these issues have been clarified in this paper via a careful micromechanical analysis of the deformation with discrete slip surfaces and dislocation structures. This analysis is supported with a functional space that allows to describe both, the continuum elastic deformation surrounding the dislocations and the displacement discontinuity induced by dislocation glide. Using extensions of the standard notions of derivative to discontinuous functions, we provide physically-sound definitions of the deformation tensors $\mathbf{F},\ \mathbf{F}^e$ and $\mathbf{F}^p$ that are purely based on kinematics. The corresponding macroscopic fields are then obtained as their continuum limit (or average) for several examples of interest, in which the multiplicative decomposition of the deformation gradient is recovered in the limit. 

The definition of $\mathbf{F}^p$ proposed in this paper is directly related
via an explicit equation (not in rate form) to the underlying slip structure
and does not refer to an artificial intermediate configuration,
but instead to a local decomposition of the plastic slip into a potentially very
large number of individual single-slip processes at the microscopic level,
which macroscopically correspond to both single slip and  simultaneous slip
over several slip systems.
As a
result, invariance requirements and plastic spin become unambiguous in the
present formulation. Additionally, the definition for $\mathbf{F}^p$ provides
a direct link to the dislocation density content in the body when referred to
the reference configuration and physically justifies previously proposed
dislocation density tensors. 

A companion mathematical paper containing a general proof for the multiplicative decomposition of the deformation gradient based on the suggested expressions for $\mathbf{F}^e$ and $\mathbf{F}^p$ is currently in progress. This proof requires energetic considerations and is out of the scope of the this work, in which the authors wanted to emphasize the kinematic character of the definitions of $\mathbf{F}^e$ and $\mathbf{F}^p$.

\appendix
\section{Incompatibility of the elastic and plastic deformation gradient at the continuum scale} \label{Sec::AppendixA}
The total deformation gradient $\mathbf{F}$ is by definition \textit{Curl}-free. However, the elastic and plastic deformation tensors ($\mathbf{F}^e$ and  $\mathbf{F}^p$) are in general incompatible, and their incompatibilities are related by
\begin{equation}
\begin{split}
&\text{Curl} \mathbf{F}^{p} = \det (\mathbf{F}) \text{curl} \mathbf{F}^{e-1} \mathbf{F}^{-T}, \quad \text{in 3D} \\
&\text{Curl} \mathbf{F}^p = \det (\mathbf{F}) \text{curl} \mathbf{F}^{e-1} , \quad \text{in 2D} \\
\end{split}
\end{equation}

The proof in three dimensions is immediate from the relation $\mathbf{F} =\mathbf{F}^e \mathbf{F}^p$
\begin{equation}
\begin{split}
\left(\text{Curl} F^p \right)_{IJ} &= F^p_{IM,K}e_{JKM} =  (F^{e-1}_{Ij} F_{jM})_{,K}e_{JKM} = F^{e-1}_{Ij,K} e_{JKM} F_{jM}\\
& = F^{e-1}_{Ij,m} e_{JKM} F_{jM} F_{mK} = F^{e-1}_{Ij,m} e_{PKM} F_{jM} F_{mK} F^{-1}_{Jn} F_{nP} \\
& = F^{e-1}_{Ij,m} e_{nmj} \det(F) F^{-T}_{nJ} = \det(F) (\text{curl} F^{e-1})_{In} F^{-T}_{nJ}\\
\end{split}
\end{equation}

Equivalently, in two dimensions, 
\begin{equation}
\begin{split} \label{Eq:CurlFeFp}
\left(\text{Curl} F^p \right)_K &= F^p_{KJ,I}  e_{3IJ}=  F^{e-1}_{Km,I} F_{mJ} e_{3IJ} = F^{e-1}_{Km,j}   F_{mJ} F_{jI} e_{3IJ}\\
& = F^{e-1}_{Km,j} \det(F) e_{3jm} = \det(F) \text{curl} F^{e-1}_K
\end{split}
\end{equation}
\section{Curl of the plastic deformation tensor}\label{Sec:AppendixB}
The distributional \textit{Curl} of a measure of the form $F^p_{IJ} = \delta_{IJ} \mathcal{L}^2 + b_I N_J \mathcal{H}^1\lfloor_{\mathcal{J}}$, where $\mathbf{b}$ is piecewise smooth over $\mathcal{J}$, is obtained by
\begin{equation}
\begin{split}
\int_{\Omega} \left(\text{Curl} F^p\right)_I \varphi \, dx &= - \int_{\Omega} F^p_{IJ} e_{KJ3} \varphi_{,K} \, dx \\
&= -\int_{\Omega \cap J} b_I N_J e_{KJ3} \left( \partial_{\mathbf{N}} \varphi N_K + \partial_{\mathbf{N}^{\bot}} \varphi N^{\bot}_K\right) \, d\mathcal{H}^1 \\
& = -\int_{\Omega \cap J} \left( b_I \partial_{\mathbf{N}} \varphi  N_J  N_K e_{KJ3}   + b_I \partial_{\mathbf{N}^{\bot}} \varphi N_J N^{\bot}_K e_{KJ3} \right) \, d\mathcal{H}^1 \\
&= \int_{\Omega \cap J} b_I \partial_{\mathbf{N}^{\bot}} \varphi \, d\mathcal{H}^1  \\
\end{split}
\end{equation} 
where $\varphi \in C_0^{\infty}$. That is, 
\begin{equation}
\text{Curl } \mathbf{F}^p = - \partial_{\mathbf{N}^{\bot}}\mathbf{b}\ \mathcal{H}^1\lfloor_{\mathcal{J}} + \llbracket \mathbf{b} \rrbracket \ \mathcal{H}^0\lfloor_{\mathcal{J}_b}
\end{equation}
where $\mathcal{J}_b$ are the points of discontinuity of the Burgers vector $\mathbf{b}$ along the jump set $\mathcal{J}$, and $ \llbracket \mathbf{b} \rrbracket$ is the jump at each of these points with the appropriate sign. This implies that only a change in the magnitude or orientation in the vector $\mathbf{b}$ induces a \textit{Curl} regardless of the shape of the jump set.

For the specific case where the jump set $\mathcal{J}$ is the union of lines (not necessarily straight) $\mathcal{J}_i$, each of which has a constant Burgers vector associated to it, $\text{Curl }\mathbf{F}^p$ is given by
\begin{equation}
\text{Curl }\mathbf{F}^p = \sum_i^{N_{\mathcal{J}}} \mathbf{b}_i \left(\delta_{\mathbf{X}_{i1}}- \delta_{\mathbf{X}_{i2}} \right),
\end{equation} 
where $\mathbf{X}_{i1}$ and $\mathbf{X}_{i2}$ are the two endpoints of the slip segment $\mathcal{J}_i$, defined such that $\left(\mathbf{X}_{i1}-\mathbf{X}_{i2}\right)\cdot \mathbf{N}_i > 0$.

\section*{Acknowledgements}

The authors acknowledge support from the Hausdorff Center for
Mathematics and the Deutsche Forschungsgemeinschaft
through  Forschergruppe 797, project CO 304/4-2.
This work performed under the auspices of the U.S. Department of
Energy by Lawrence Livermore National Laboratory under Contract
DE-AC52-07NA27344. 

\bibliographystyle{plainnat}
\bibliography{Biblio}

\end{document}